\documentclass[
    referee,
    linenumbers,
    iicol,
    oneside
]{sn-jnl}

\usepackage{amsmath,amssymb,amsfonts,amsthm}
\usepackage{graphicx}
\usepackage{booktabs}
\usepackage{siunitx}
\usepackage{url}
\usepackage[utf8]{inputenc}
\usepackage{algorithm}
\usepackage{algorithmic}
\usepackage{caption}
\usepackage{float}
\usepackage{subcaption}
\usepackage{xcolor} 
\usepackage{mathtools}
\usepackage{bm}

\usepackage[super,square]{natbib}

\usepackage{hyperref}
\hypersetup{colorlinks=true, citecolor=blue, urlcolor=blue, linkcolor=blue, pdfborder={0 0 0}}

\title{Solving Large-Scale Parameter Inference in Controllable Nuclear Fusion via Coherent Ising Machines and Block Coordinate Descent}

\author[1]{Zijie Liu}
\author[2]{Wenchao Liu}
\author[1,3]{Yao Huang\thanks{*Corresponding author email: yaohuang@ipp.ac.cn}}
\author[2]{Kai Wen\thanks{*Corresponding author email: contact@shturl.}}
\author[1,3]{Yuehang Wang\thanks{*Corresponding author email: wangyhplasma@ipp.ac.cn}}
\author[2]{Wenxin Li}
\author[1,3]{Zhengping Luo}
\author[2]{Anzhe Jiang}
\author[1,3]{Wangyi Rui}
\author[2]{Hongdong Zhu}
\author[1,3]{Kai Wu}
\author[2]{Qi Gao}
\author[1,3]{Bihao Guo}
\author[2]{Jingjing Wang}
\author[1,3]{Shuai Li}
\author[2]{Yin Ma}
\author[1,3]{Dalong Chen}
\author[2]{Hai Wei}
\author[1,3]{Qiping Yuan}
\author[1,3]{Bingjia Xiao}
\author[3]{Jiangang Li}

\affil[1]{Institute of Energy, Hefei Comprehensive National Science Center, Anhui, Hefei, 230031, China}
\affil[2]{Beijing QBoson Quantum Technology Co., Ltd, Beijing 100015, China}
\affil[3]{Institute of Plasma Physics, Hefei Institutes of Physical Science, Chinese Academy of Sciences, Hefei 230031, China}

\abstract{
Bayesian inference offers significant advantages in tokamak integrated data analysis, enabling the fusion of multi-source diagnostic information and achieving self-consistent inversion of plasma parameters. However, when the likelihood probability based on diagnostics follows a nonlinear Gaussian distribution, the parameters corresponding to the maximum of the posterior distribution are generally solved using sampling methods. This procedure incurs heavy computational overhead and struggle to satisfy the requirements of real-time plasma control. To address this bottleneck, this paper proposes to realize real-time Bayesian solution of plasma parameters utilizing the Coherent Ising Machine (CIM) which is quantum computer. Based on coupled degenerate optical parametric oscillators (DOPOs), it possesses distinctive advantages in addressing high-dimensional parameter inference tasks. Natively equipped with arbitrary all-to-all connectivity, CIM enables the direct mapping of the negative log-Gaussian posterior potential onto the spin Hamiltonian, and leverages quantum vacuum fluctuations to perform rapid state-space search during pitchfork bifurcation evolution. Taking plasma current profile inversion on the EAST device as a validation case, this work employs quantum computing to solve the posterior probability within Bayesian inference. The feasibility of this scheme is verified, and the solution time for a single run reaches approximately 1 millisecond. Restricted by the qubit count of existing quantum computers, the solving procedure has to adopt a block-wise iterative approach at the present stage, which prevents it from satisfying the performance requirements for real-time computation. With the continuous advancement of quantum hardware, the qubit scale of quantum computers is growing exponentially. In the future, real-time inversion of plasma parameters is fully expected to be realized, providing efficient data support for the operation and control of tokamak devices.
}
\keywords{Bayesian inference; Plasma parameters; Coherent Ising machine; Quantum computer; Fusion}

\begin{document}
\maketitle 

\makeatletter
\clubpenalty=0
\widowpenalty=0
\interlinepenalty=0
\makeatother

\vspace{-32pt}

\section{Introduction}
Sustained and stable confinement of high-temperature plasma constitutes an essential prerequisite for achieving controlled nuclear fusion\cite{doyle2007confinement,wagner2007high,han2022sustained}. High-performance plasma operation demands optimized regulation of plasma profile parameters, such as the plasma current density profile, electron density profile, and electron temperature profile\cite{argomedo2013lyapunov,Barton2015,kim2026realtime}. To acquire internal information on these plasma profile quantities, a comprehensive suite of diagnostic systems has been deployed on tokamak devices\cite{boivin2005diag,lan2022cnn}. These multi-angle, multi-position diagnostic systems monitor in-vessel plasma parameters and deliver reliable datasets for tokamak experiments and physical interpretation\cite{jalal2025multimodal,xie2025nn,denk2025simultaneous}. Nevertheless, profile reconstructions from separate diagnostics frequently suffer from inconsistency caused by variations in measurement geometry, noise features and reconstruction methods, hence integrated data analysis is essential for retrieving reliable, self-consistent plasma parameter profiles\cite{liu2024plasma,Rutigliano_2026}.

The concept of IDA (integrated data analysis) first emerged in 1985, which aims to comprehensively analyze data related to the same physical parameters to obtain more reliable results \cite{CHRISTIANSEN198785}. IDA has been widely applied in fusion research, and a variety of IDA programs have been developed, such as  MINERVA \cite{svensson2007large}, OMFIT \cite{Orso_MENEGHINI2013}, PTRANSP \cite{Kritz2009}, etc. Compared to other IDA programs, MINERVA is based on Bayesian inference and facilitates the integration of diagnostics through joint probability. It constructs the likelihood probability based on the forward physical models of each diagnostic and constrains the parameter profiles via prior probabilities. MINERVA integrates the likelihood probabilities derived from multiple diagnostics together with prior probability into the joint posterior probability. It thereby calculates self-consistent plasma parameter profiles and quantifies the uncertainties of these profile parameters. IDA based on Bayesian inference was initially proposed by Fischer $et$ $al.$ for the W7-AS device, and it was used for the integrated inversion of plasma electron density and temperature profiles \cite{Fischer_2002,Fischer_2003}. MINERVA was subsequently developed based on this. Integrated diagnostic inversion was also achieved through IDA based on Bayesian inference for other fusion devices, such as TJ-II\cite{2011Integrated}, ASDEX Upgrade \cite{dfischer2010}, JET \cite{svensson2008current, kwak2022bayesian,Schmuck_2020,Kwak_2020}, MAST \cite{hole2010use,von2014recent}, HL-2A \cite{PAN_2022,Chen_2023,zjliu2023}, EAST \cite{liu2022advanced,liu2021plasma,Liu_2022,liu2024electrondensityEAST}, etc.

Although IDA based on Bayesian inference can resolve inconsistencies across multi-diagnostics, sampling-based solvers such as Markov Chain Monte Carlo (MCMC) are mandatory to compute posterior distributions when nonlinear forward models prohibit the construction of linear Gaussian likelihoods. For ultra-high-dimensional posterior problems arising from tokamak plasma profile reconstruction, classical sampling methods incur prohibitive computational latency, with single reconstruction runs consuming tens of hours or even multiple days—far beyond the millisecond timescale required for real-time plasma feedback control.
To overcome this critical real-time bottleneck, this work leverages coherent optical computing via {\color{blue}Coherent Ising Machines (CIMs)~\cite{Inagaki2016CIM, Wei2026QBosonCIM}} to accelerate high-dimensional Bayesian parameter inference. By exploiting quantum vacuum fluctuations to seed the gain-competition dynamics during the bifurcation process, CIMs can parallelly explore the high-dimensional energy landscape and rapidly identify the ground state. We reformulate the Bayesian inference objective into quadratic unconstrained binary optimization (QUBO) and Ising Hamiltonian forms natively compatible with the analog optical solver. This continuous-to-binary mapping and QUBO reformulation constitutes a general-purpose methodology, offering a universal mathematical interface to map arbitrary continuous physical inverse problems onto discrete analog Ising solvers. A domain-decomposed block coordinate descent (BCD) strategy further partitions the global large-scale optimization task to mitigate current limitations on the spin capacity of analog Ising hardware. As quantum and optical technologies advance towards ultra-large-scale spin capacities (supporting $10^5$ to $10^6$ spins), BCD partitioning will become obsolete, enabling a single-step global optimization that reduces the solver latency to the physical bifurcation scale of the CIM, thereby realizing ultimate real-time active feedback control.

This coherent-classical hybrid BCD-CIM paradigm constitutes the core contribution of this paper and delivers a feasible route toward real-time closed-loop control of tokamak plasmas. Specifically, given the physical and hardware constraints of current devices, this study is positioned as a proof-of-principle algorithm validation. We validate this hybrid BCD-CIM paradigm on the specific physical scenario of plasma current profile reconstruction \cite{liu2021plasma}. Taking the Bayesian-inference-based inversion of plasma current density profiles as the validation testbed offers the following merits: \\
1) Integrated data analysis grounded in Bayesian inference typically adopts Gaussian distributions; hence, plasma current density profile inversion serves as a representative validation scenario. \\
2) The inference of plasma current density relies on electromagnetic measurements, whose likelihood obeys a linear Gaussian distribution. This guarantees a unique, deterministic maximum a posteriori(MAP) solution, which acts as a rigorous and exclusive mathematical benchmark to verify the correctness and convergence of the CIM-based solver. \\
3) Plasma current density constitutes a fundamental plasma profile parameter. Inversions of other plasma profile quantities, such as electron density and electron temperature profiles, all depend on the magnetic flux profiles derived from plasma current calculations. \\
It is anticipated that the hybrid framework proposed in this work can cut reconstruction latency from tens of hours down to the millisecond regime, accurately resolve fine plasma structures, and deliver computational speedups of multiple orders of magnitude.

\section{Results}\label{sec:results}

The reconstruction of plasma parameter profile represents a pivotal inverse problem in magnetic confinement fusion, fundamentally limited by the computational overhead of high-dimensional optimization. To overcome this bottleneck, we develop and systematically evaluate a novel parallel optimization framework, leveraging the computational paradigm of Coherent Ising Machines (CIM), against traditional iterative schemes including the {\color{black} Fast Simulated Annealing (FSA)} and varying acceleration strategies. Our experimental validation investigates four critical aspects of the algorithmic design space: convergence efficacy of parallel frameworks, spatial partitioning strategies, the impact of Nesterov-type (FISTA) acceleration~\cite{beck2009fast}, and the refinement capability of a two-stage hybrid optimization scheme.

{\color{black} In addition to these hardware-constrained iterative configurations, we first investigate the target operating regime of the proposed CIM framework, in which sufficient spin capacity allows the entire plasma current inference problem to be encoded and solved globally without spatial decomposition. The results are therefore organized into two complementary regimes: a one-step global solution under sufficient CIM capacity and a block-wise iterative solution under the spin-capacity constraints of currently available hardware.}

{\color{black}

\subsection{One-Step Global Solution with Sufficient CIM Capacity}

The monolithic plasma current reconstruction problem can be directly encoded into a QUBO problem. When sufficient spin resources are available, this monolithic QUBO can be solved globally in one step by the CIM without invoking the block decomposition framework.

The primary bottleneck for the one-step full-scale plasma current reconstruction problem is the number of available spins. For $N$ continuous current variables represented by $K$ binary bits each, the required number of spins scales as $O(NK)$. The full-scale reconstruction problem considered in this work requires approximately $2{,}039$ spins, which exceeds the effective spin capacity of existing CIM hardware.

To verify the feasibility of the one-step global solution, we evaluated two coarser plasma current reconstruction problems, containing 33 and 94 current variables, that fit directly into a 550-spin CIM. For each problem, all variables were simultaneously encoded into a monolithic QUBO problem and solved directly by the CIM. Two-stage adaptive binarization with $K_{\rm coarse}=10$ and $K_{\rm fine}=5$ (see \ref{subsubsec:two_stage_binarization}) was adopted for high-precision encoding. The resulting reconstructions were evaluated against the deterministic MAP reference solutions obtained from the corresponding quadratic Bayesian inference problems.

For both problems, the one-step CIM accurately reconstructed the reference current distributions within milliseconds (Fig.~\ref{fig:one_step_solution}). The relative $L_2$ errors of the corresponding reconstructions were $0.040$ and $0.039$, with relative objective gaps of $1.6\times 10^{-5}$ and $7.3\times 10^{-5}$, respectively.

Fig.~\ref{fig:one_step_runtime} illustrates the reliability and computational latency of the one-step solution across 100 independent runs for each problem. For the 33- and 94-variable models, the median CIM solution times were $4.36$ms and $4.67$ms, with standard deviations of $4.88$ms and $3.87$ms, respectively. 

These results demonstrate that, once a monolithic problem fits within the CIM spin capacity, it can be solved efficiently and reliably by the one-step CIM. As the capacity of optical CIM systems continues to scale (Fig.~\ref{fig:cim_scaling}), increasingly large plasma current reconstruction problems can be accommodated and solved directly within this one-step framework.

\begin{figure*}[htbp]
    \centering
    \begin{subfigure}{0.7\textwidth}
        \includegraphics[width=\linewidth]{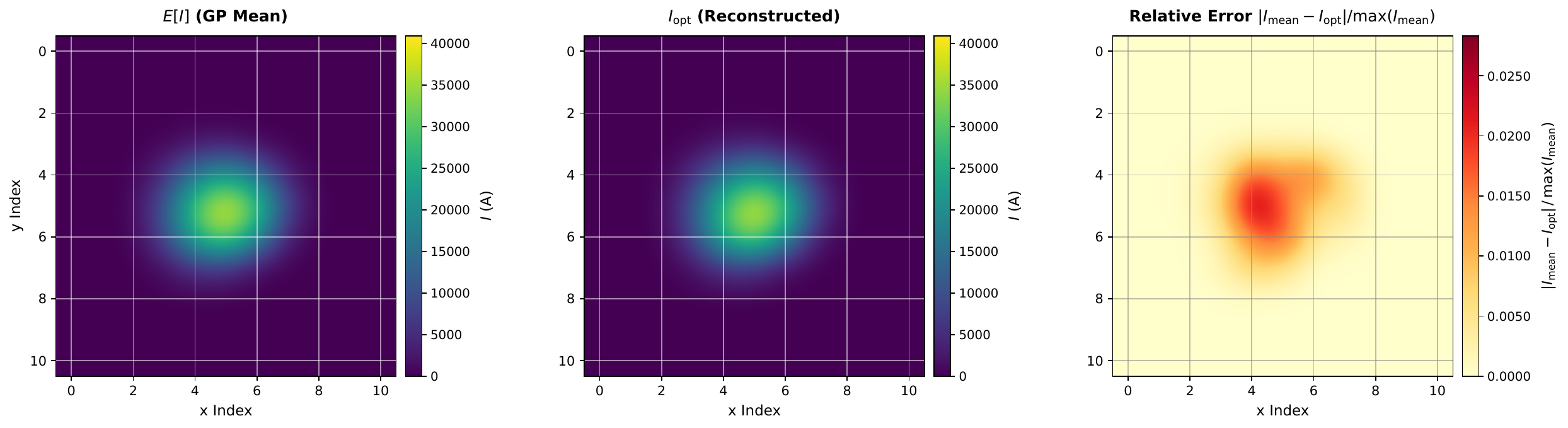}
        \caption{33-variable Model}
        \label{fig:one_step_n_actv_33}
    \end{subfigure}
    \\
    \begin{subfigure}{0.7\textwidth}
        \includegraphics[width=\linewidth]{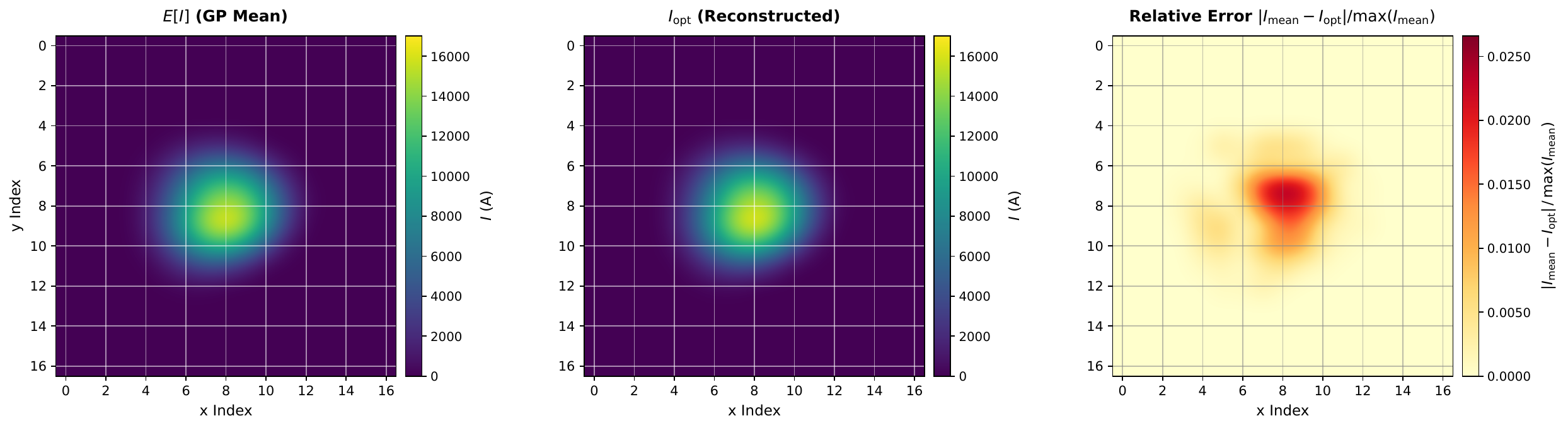}
        \caption{94-variable Model}
        \label{fig:one_step_n_actv_94}
    \end{subfigure}
    \\
    \begin{subfigure}{0.7\textwidth}
        \includegraphics[width=\linewidth]{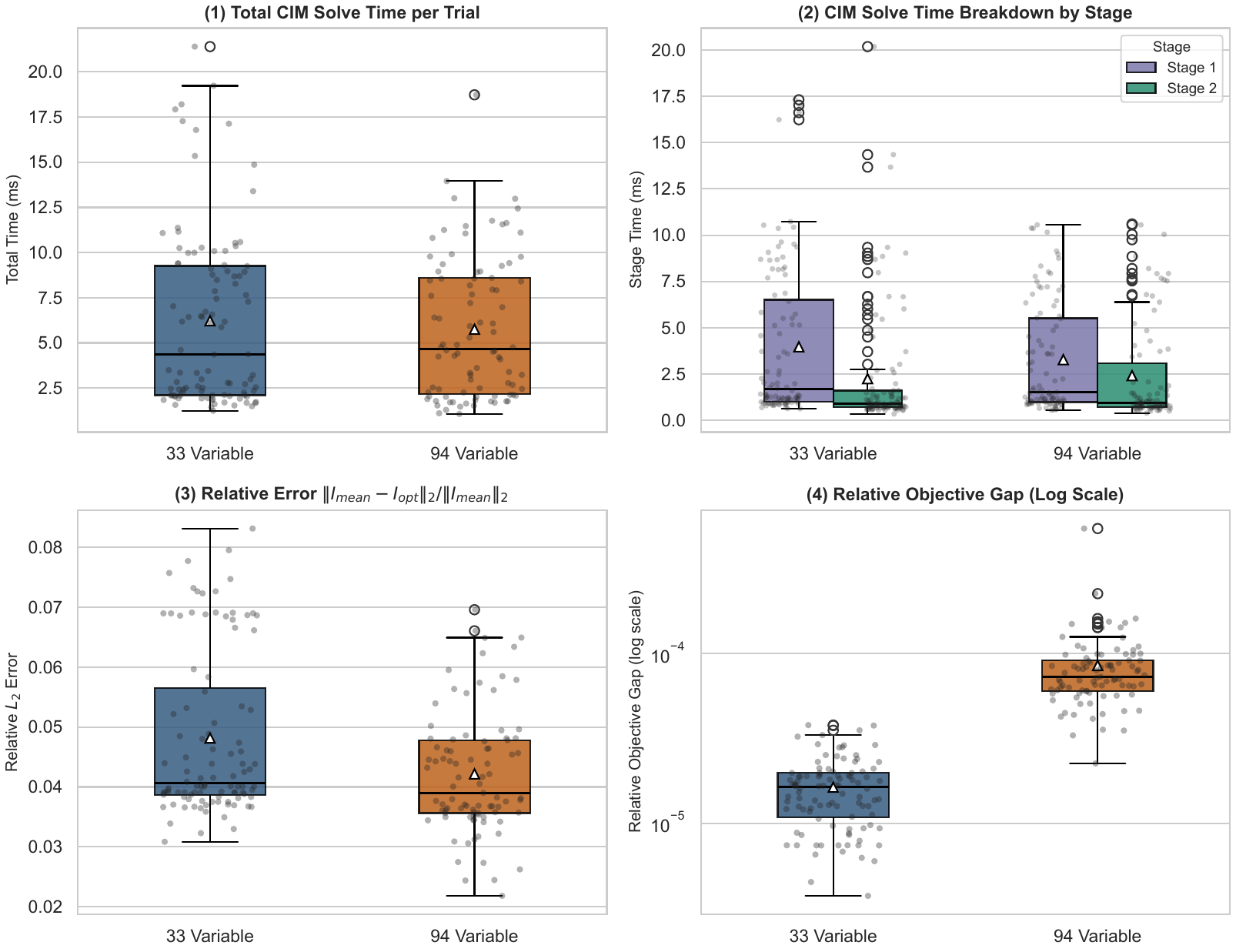}
        \caption{Solving Time and Quality}
        \label{fig:one_step_runtime}
    \end{subfigure}

    \caption{One-step global CIM solution under sufficient spin capacity.
    (a--b) Current reconstruction results for the 33- and 94-variable models.
    (c) CIM solving-time and quality across 100 independent runs.}
    \label{fig:one_step_solution}
\end{figure*}

\begin{figure*}[htbp]
    \centering
    \includegraphics[width=0.70\textwidth]{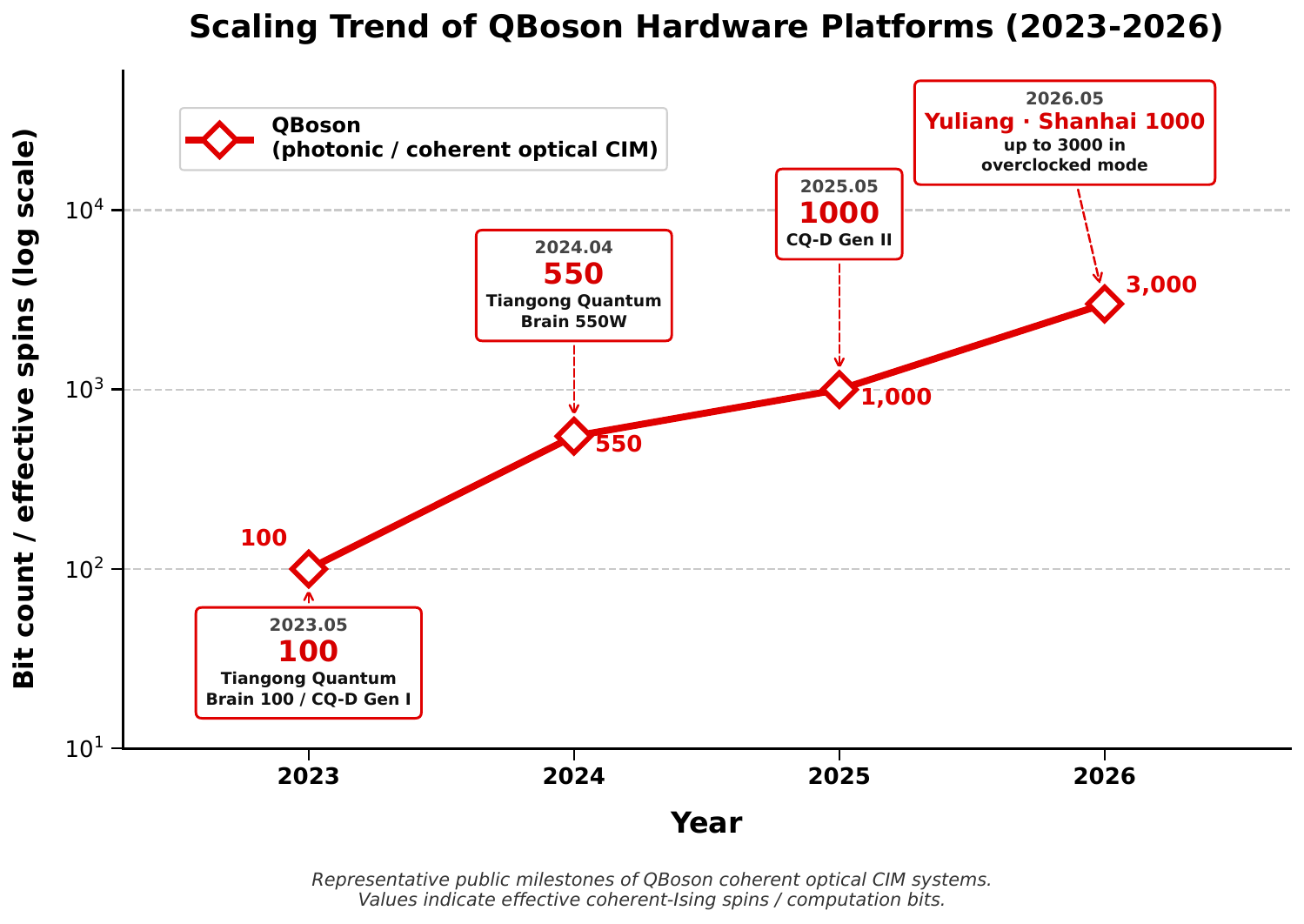}
    \caption{Evolution of the implemented computational scale of CIM systems and comparison with the spin requirements for one-step plasma current reconstruction.}
    \label{fig:cim_scaling}
\end{figure*}

}

{\color{black}

\subsection{Block-Wise Iterative Solution under Current Hardware Constraints}
At present, the spin requirements of the full-scale current reconstruction problem exceed the capacity of the available CIM hardware. To retain the capability of applying CIM optimization to the original large-scale inference problem, we therefore employ the domain-decomposed BCD-CIM framework as a practical hardware-compatible alternative. The following experiments evaluate the convergence, spatial partitioning, acceleration, and refinement properties of this block-wise iterative strategy.

}

\subsubsection{Superiority of CIM-Based Parallel Iteration}

\begin{figure*}[htbp]
    \centering
    \begin{subfigure}{0.32\textwidth}
        \includegraphics[width=\linewidth]{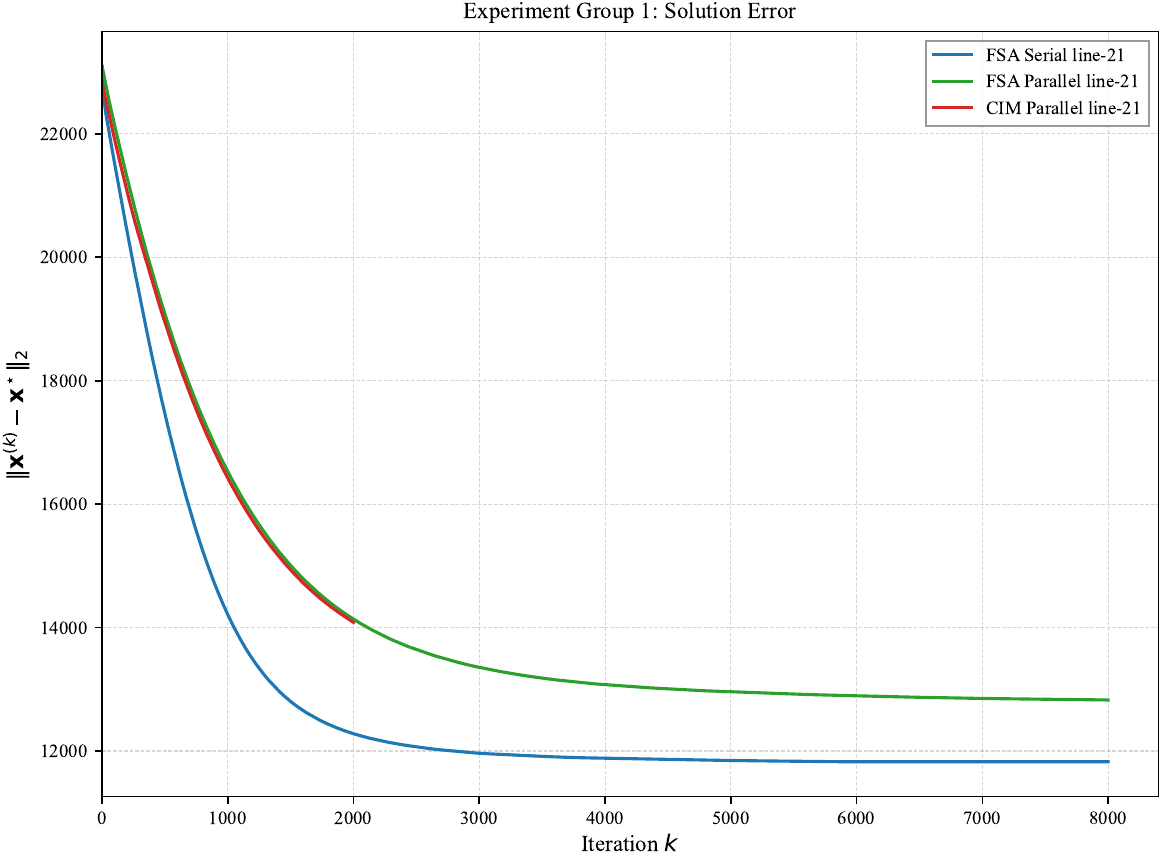}
        \caption{Error Convergence}
        \label{fig:exp1_error}
    \end{subfigure}
    \hfill
    \begin{subfigure}{0.32\textwidth}
        \includegraphics[width=\linewidth]{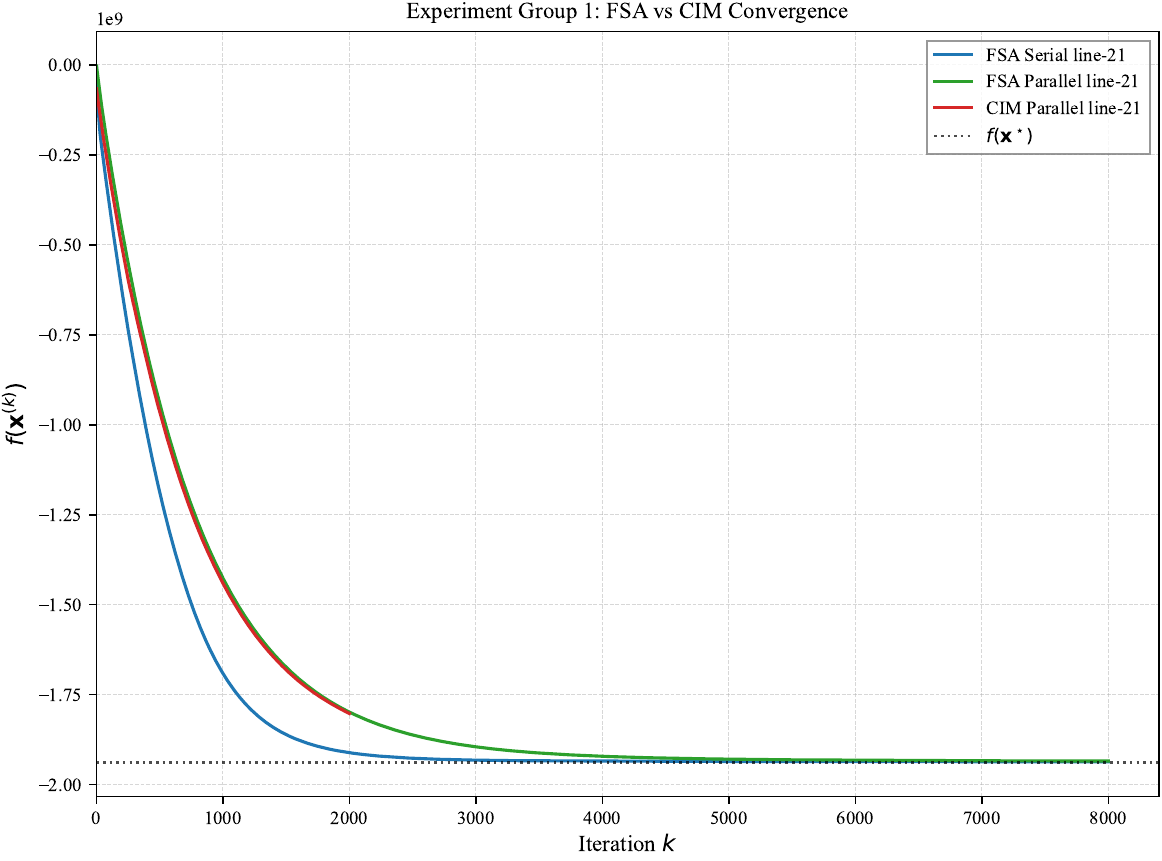}
        \caption{Objective Function}
        \label{fig:exp1_fx}
    \end{subfigure}
    \hfill
    \begin{subfigure}{0.32\textwidth}
        \includegraphics[width=\linewidth]{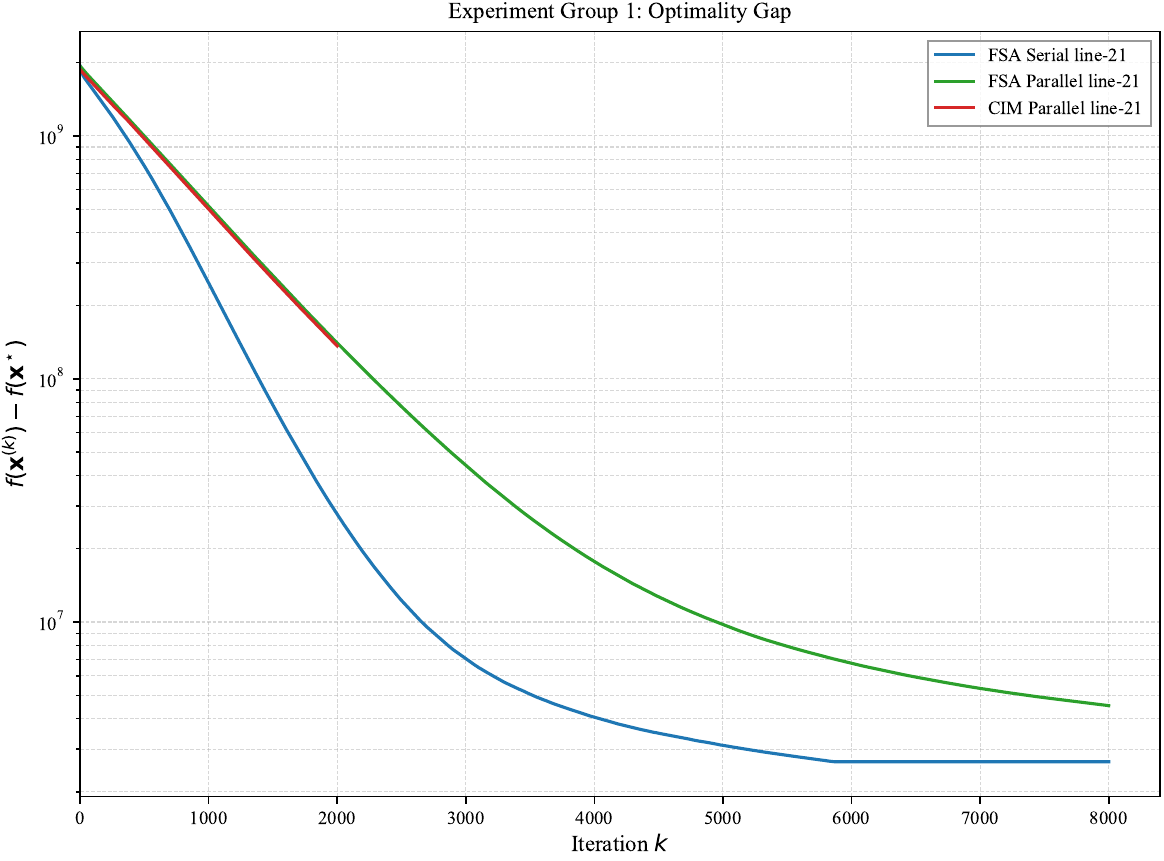}
        \caption{Objective Gap}
        \label{fig:exp1_gap}
    \end{subfigure}
    \caption{Convergence profiles comparing FSA (serial and parallel) with the CIM parallel framework.}
    \label{fig:exp1_convergence}
\end{figure*}

\begin{figure*}[htbp]
    \centering
    \includegraphics[width=0.7\textwidth]{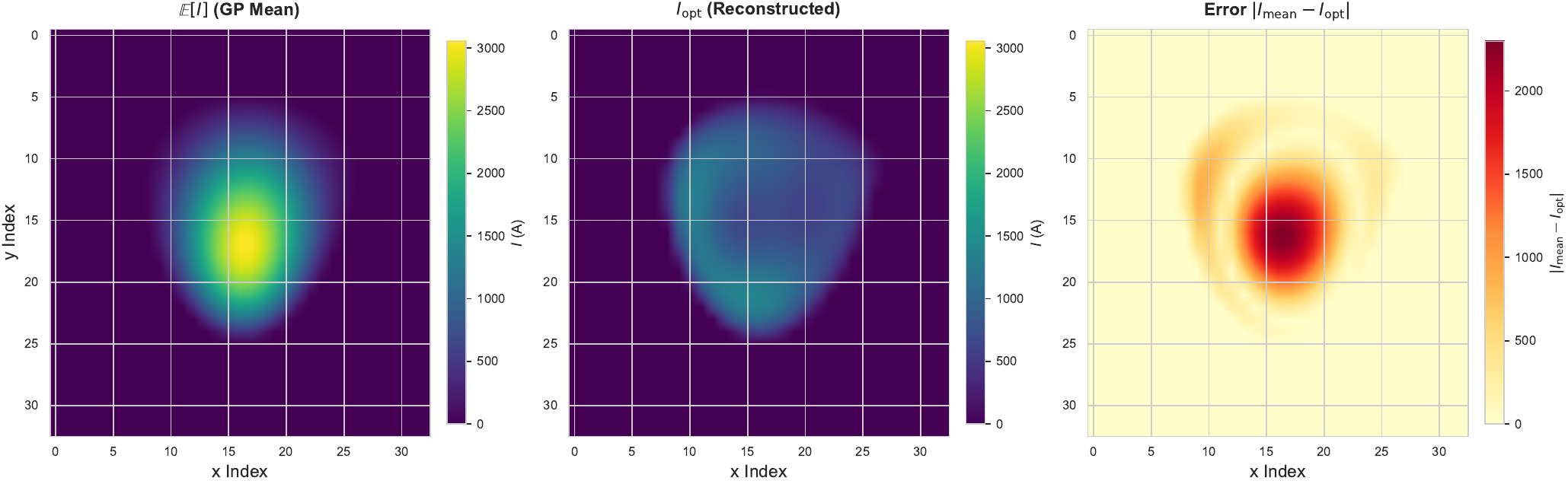}
    \caption{Current density distribution reconstructed by CIM parallel iteration at 2000 iterations.}
    \label{fig:exp1_current}
\end{figure*}

To establish a performance baseline for large-scale current density reconstruction, we first evaluated the algorithmic convergence and reconstruction fidelity of an unaccelerated CIM parallel framework against both serial and parallel implementations of the standard FSA approach (utilizing identical band-matrix spatial partitioning). As illustrated by the convergence trajectories of the reconstruction error (Fig.~\ref{fig:exp1_error}) and target function mismatch (Fig.~\ref{fig:exp1_fx}), the CIM parallel approach demonstrates a radically accelerated convergence profile. Notably, the CIM framework required merely 2,000 iterations to achieve a substantially lower reconstruction error and tighter objective gap (Fig.~\ref{fig:exp1_gap}) compared to the FSA configurations running for 8,000 iterations. The final current density distribution reconstructed by CIM (Fig.~\ref{fig:exp1_current}) perfectly captures the desired spatial topology and reference modes, corroborating that CIM-inspired massive parallelism fundamentally breaks the scaling limits of conventional sequential evaluation in fusion applications. {\color{black} Here, the GP mean denotes the posterior mean of the Gaussian Bayesian model, which coincides with the unique MAP solution and is used as the reference solution.}

\subsubsection{Spatial Partitioning and Algorithmic Acceleration}

\begin{figure*}[htbp]
    \centering
    \begin{subfigure}{0.32\textwidth}
        \includegraphics[width=\linewidth]{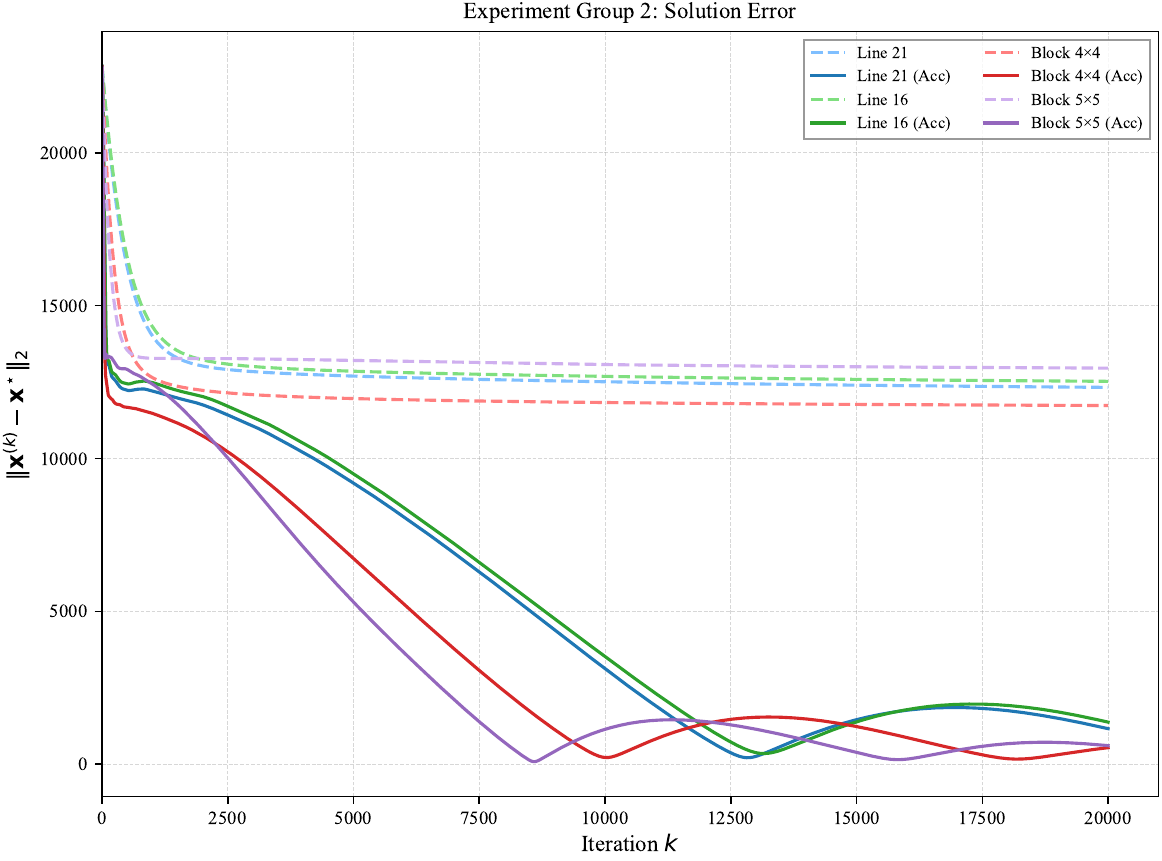}
        \caption{Error (Group 2)}
    \end{subfigure}
    \hfill
    \begin{subfigure}{0.32\textwidth}
        \includegraphics[width=\linewidth]{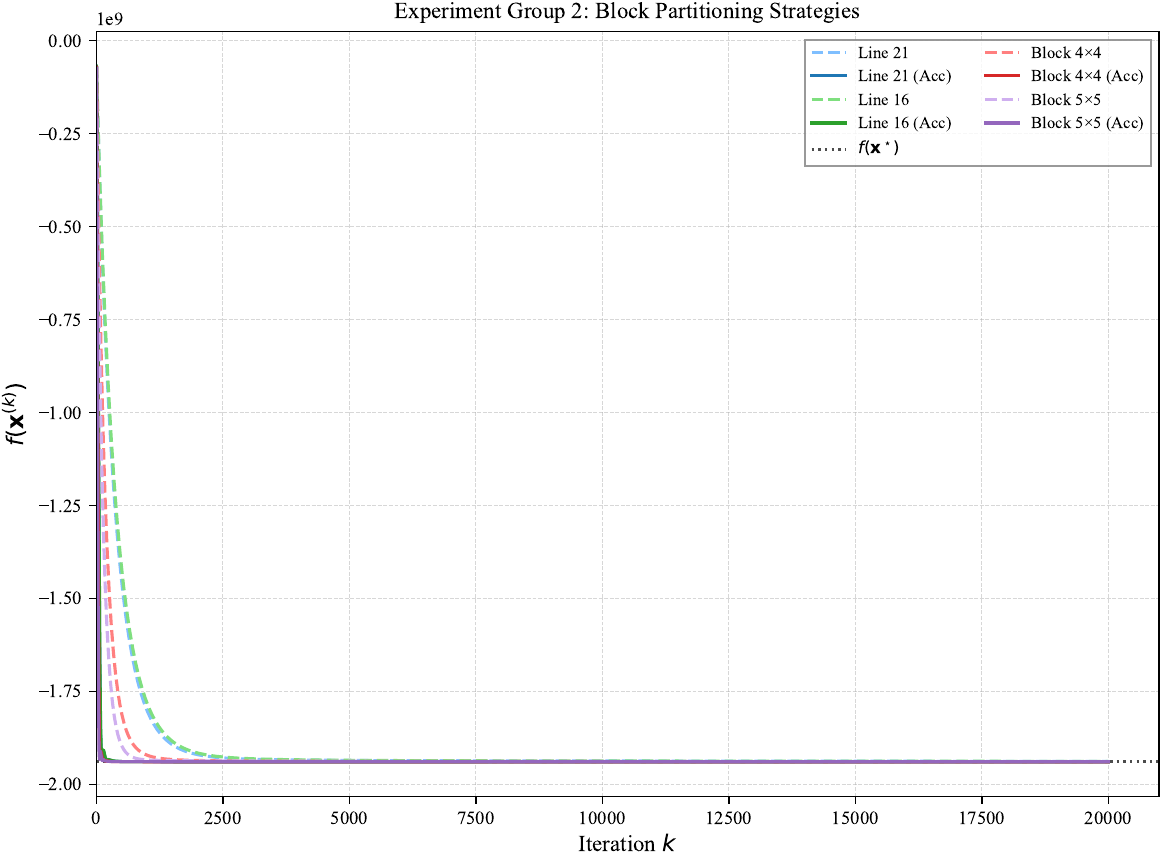}
        \caption{Objective (Group 2)}
    \end{subfigure}
    \hfill
    \begin{subfigure}{0.32\textwidth}
        \includegraphics[width=\linewidth]{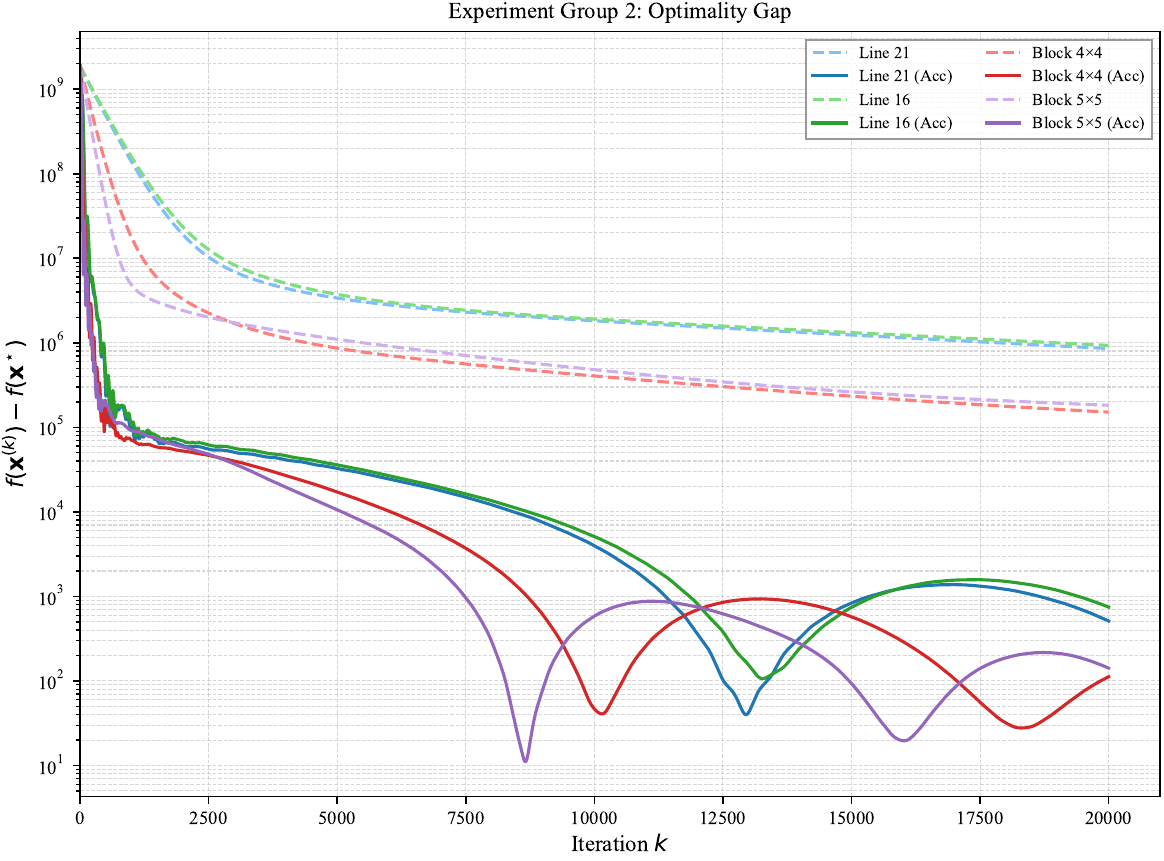}
        \caption{Gap (Group 2)}
    \end{subfigure}
    \caption{Effects of spatial partitioning architectures and FISTA acceleration on serial iterative convergence.}
    \label{fig:exp2_convergence}
\end{figure*}

The efficiency of any block coordinate descent or splitting method is mathematically coupled with its spatial partitioning topology. To systematically decouple this effect, we assessed serial analytical iterations across both band-wise configurations (bandwidths of 16 and 21) and localized block-wise schemas ($4 \times 4$ and $5 \times 5$ grids), tracking evolution up to 20,000 iterations (Fig.~\ref{fig:exp2_convergence}). Furthermore, each structural configuration was evaluated with and without the Fast Iterative Shrinkage-Thresholding Algorithm (FISTA) momentum scheme. 

\begin{figure*}[htbp]
    \centering
    \includegraphics[width=0.7\textwidth]{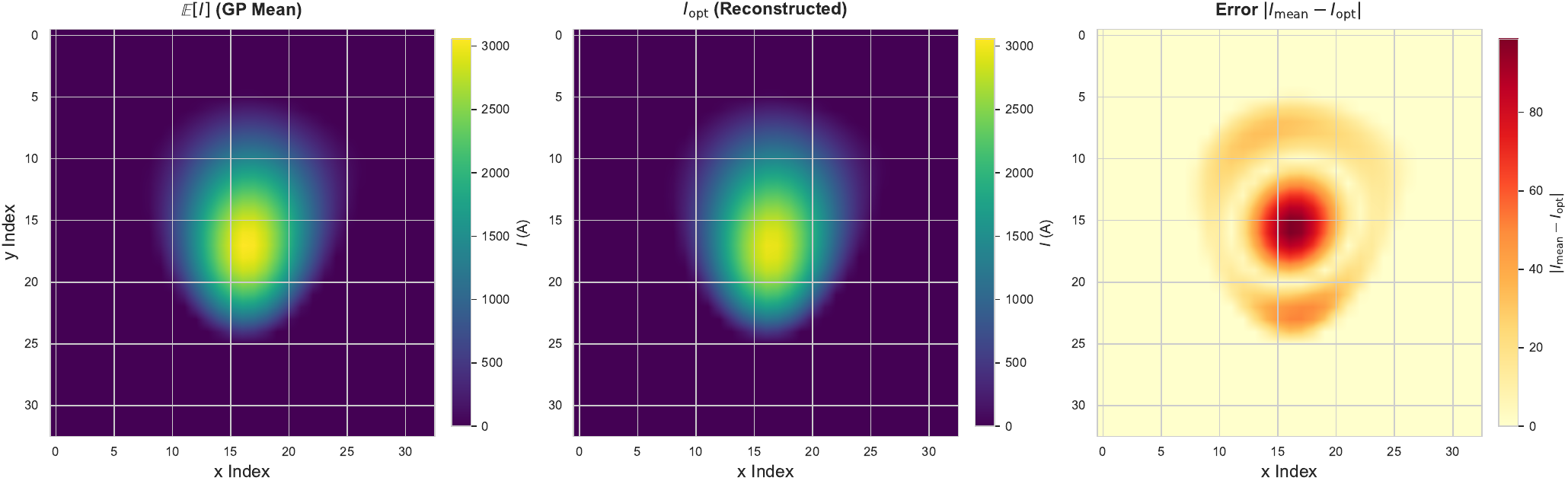}
    \caption{Current density reconstruction using $4 \times 4$ block partitioning with FISTA acceleration.}
    \label{fig:exp2_current}
\end{figure*}

\begin{figure*}[htbp]
    \centering
    \begin{subfigure}{0.32\textwidth}
        \includegraphics[width=\linewidth]{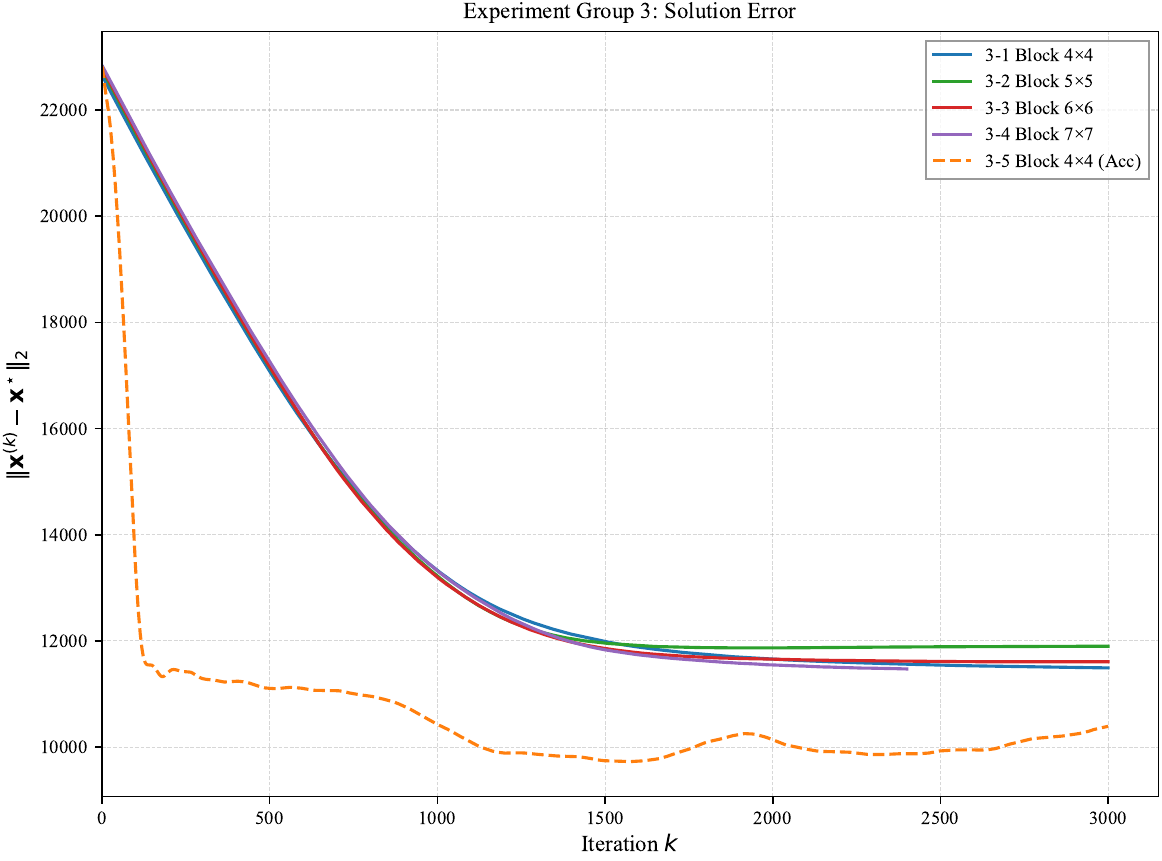}
        \caption{Error (Group 3)}
    \end{subfigure}
    \hfill
    \begin{subfigure}{0.32\textwidth}
        \includegraphics[width=\linewidth]{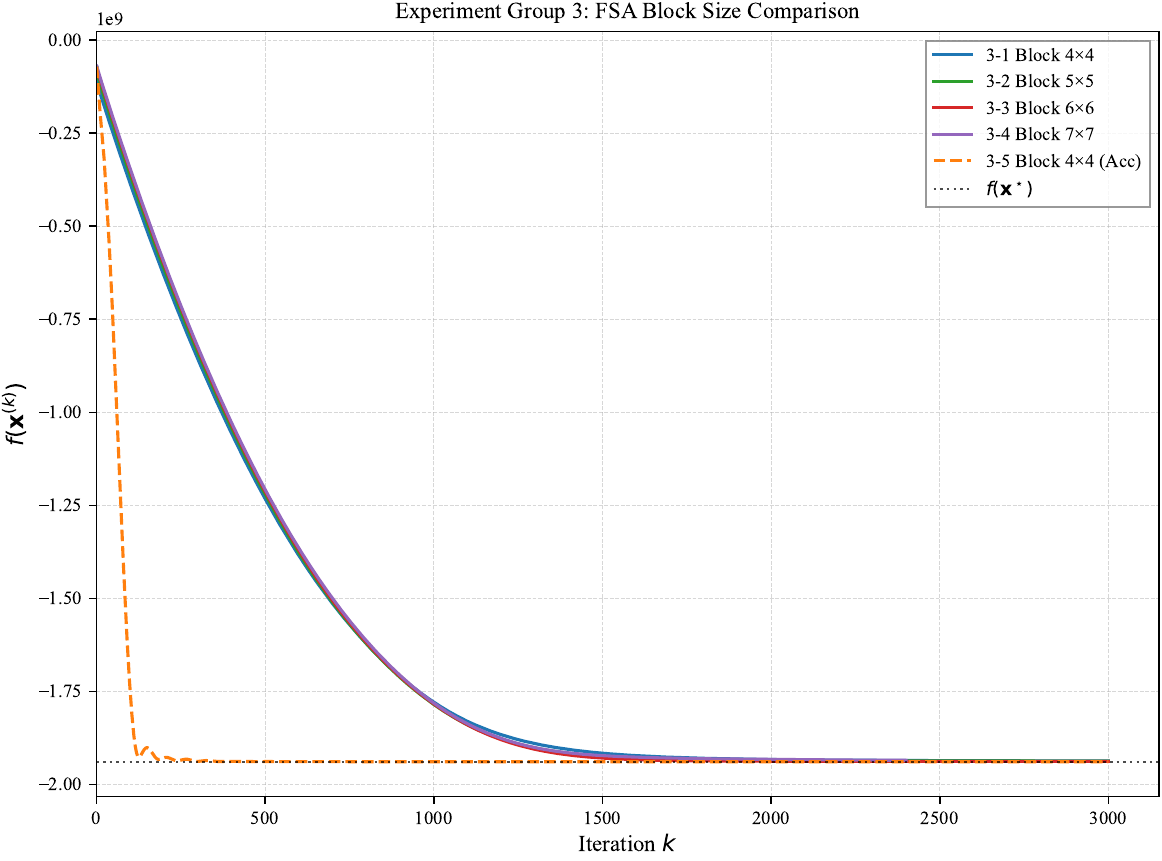}
        \caption{Objective (Group 3)}
    \end{subfigure}
    \hfill
    \begin{subfigure}{0.32\textwidth}
        \includegraphics[width=\linewidth]{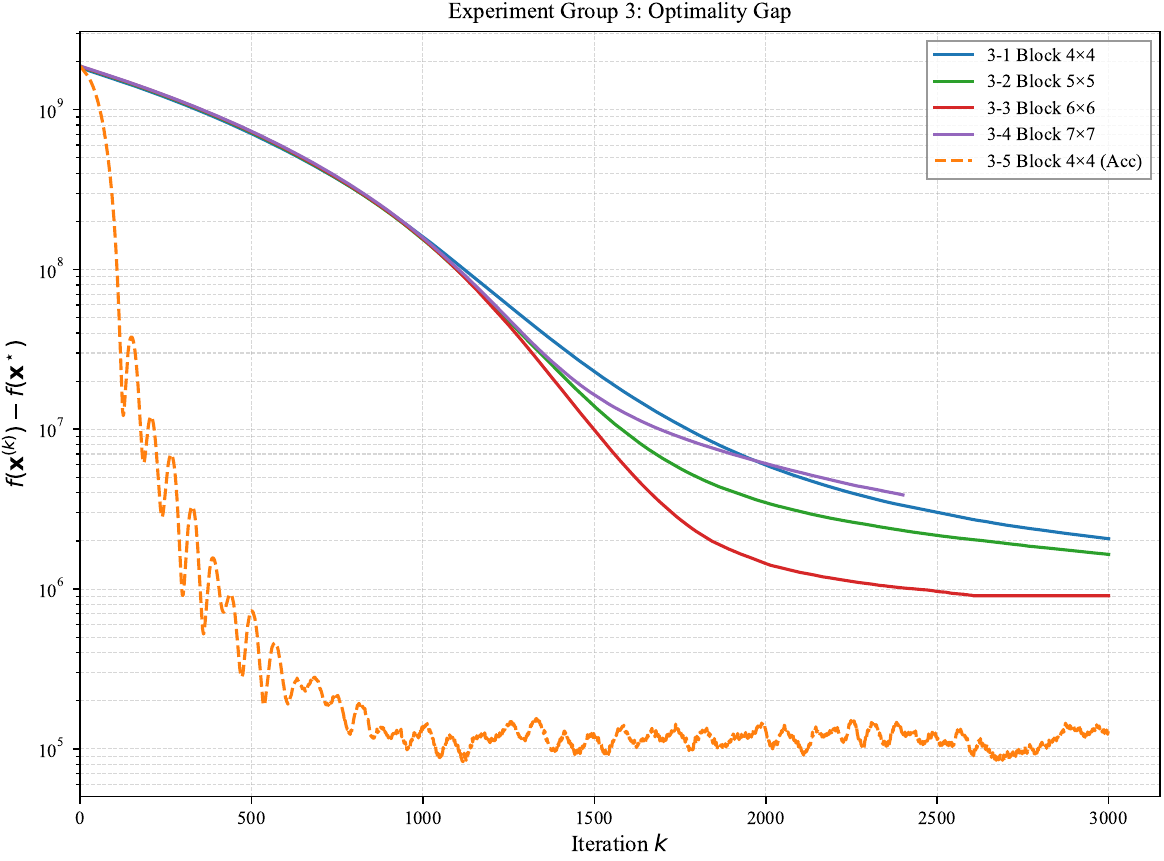}
        \caption{Gap (Group 3)}
    \end{subfigure}
    \caption{Convergence behavior for varying block sizes executing FSA serial iterations.}
    \label{fig:exp3_convergence}
\end{figure*}

Our findings indicate a strong geometric dependency in the optimization landscape: structurally compact block-wise partitions functionally isolate independent reconstruction domains, mitigating long-range interference errors inherent in wide band-matrices. Specifically, the $4 \times 4$ block partition configuration exhibited optimal intrinsic convergence. When coupled with FISTA acceleration, this configuration yielded the steepest descent gradient in both the primary objective function and error metrics, consistently bypassing the local minima trap that slows down unaccelerated iterations. The high fidelity of this approach is visualized in Fig.~\ref{fig:exp2_current}. Further parametric studies on the FSA framework (Fig.~\ref{fig:exp3_convergence}) corroborated this scaling behavior: as block size varies from $4 \times 4$ to $7 \times 7$, the optimal balance of intra-block coupling and inter-block decoupling is consistently localized around the $4 \times 4$ grid, which---enhanced by FISTA---reduces computational effort by orders of magnitude while preserving fine-grained plasma characteristic features.

\subsubsection{High-Fidelity Refinement via Hybrid Two-Stage Optimization}

While primary iterative techniques efficiently collapse the global macroscopic error, they typically struggle with microscopic high-frequency artifacts (residual ``ripples'') near the convergence floor due to the discrete nature of binary encoding. To resolve this, we introduced a two-stage hybrid optimization topology: a primary global sweep (utilizing 2,000 algorithmic cycles of varying architectures such as accelerated FSA and unaccelerated CIM) seamlessly coupled with a Conjugate Gradient (CG) continuation phase ($\approx 150$ iterations) for asymptotic refinement in the continuous domain.

\begin{figure*}[htbp]
    \centering
    \begin{subfigure}{0.32\textwidth}
        \includegraphics[width=\linewidth]{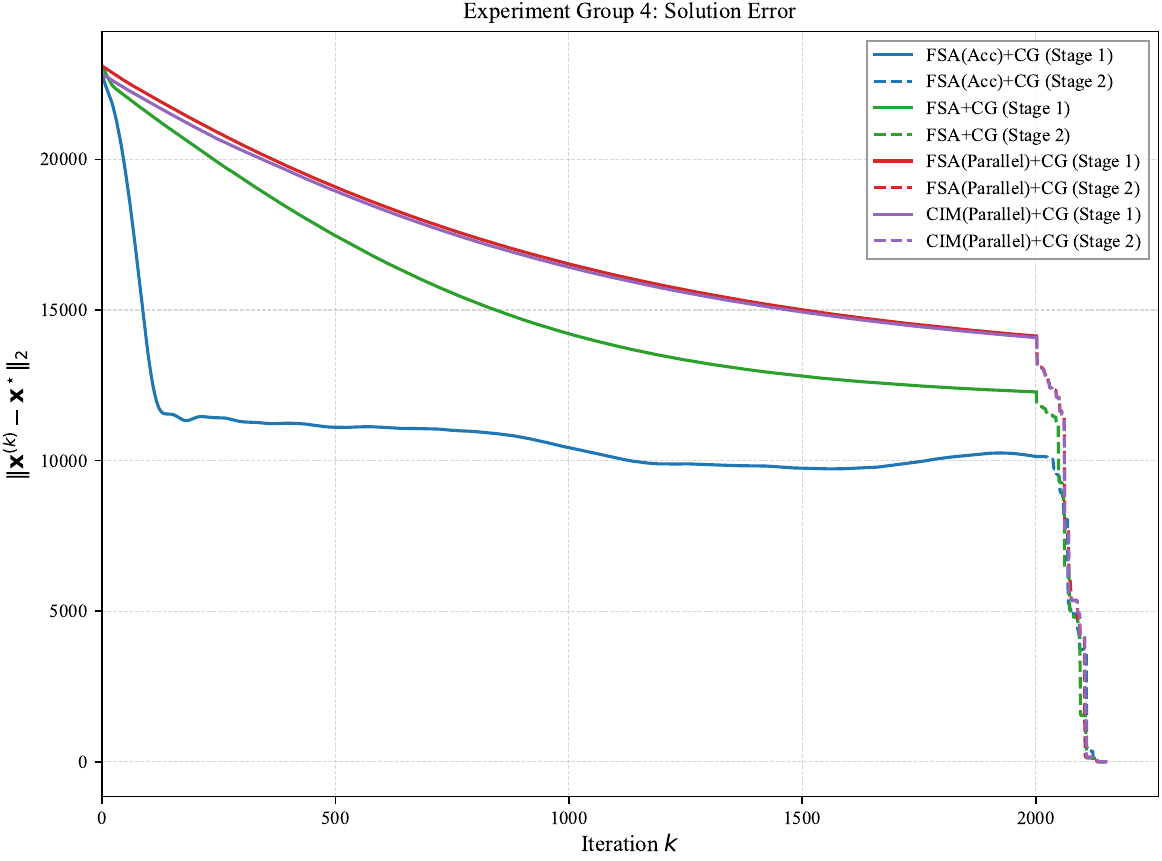}
        \caption{Error (Group 4)}
    \end{subfigure}
    \hfill
    \begin{subfigure}{0.32\textwidth}
        \includegraphics[width=\linewidth]{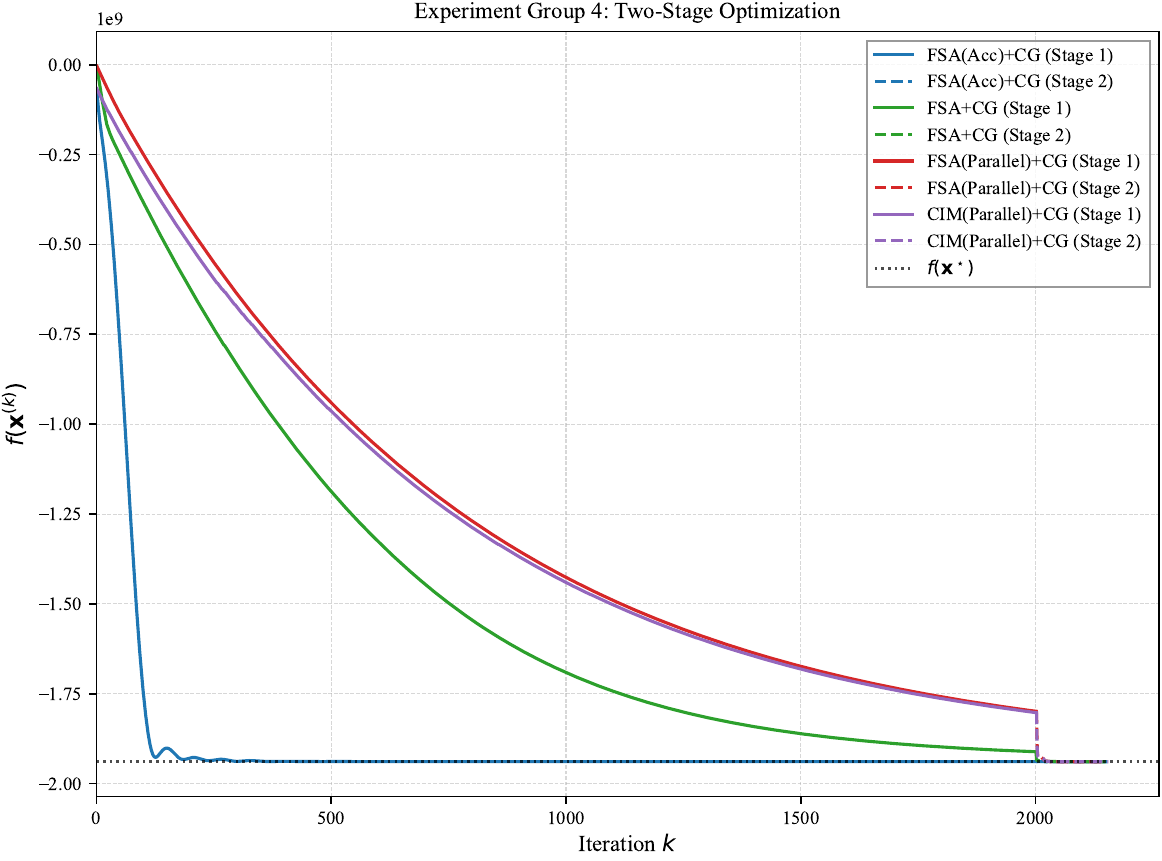}
        \caption{Objective (Group 4)}
    \end{subfigure}
    \hfill
    \begin{subfigure}{0.32\textwidth}
        \includegraphics[width=\linewidth]{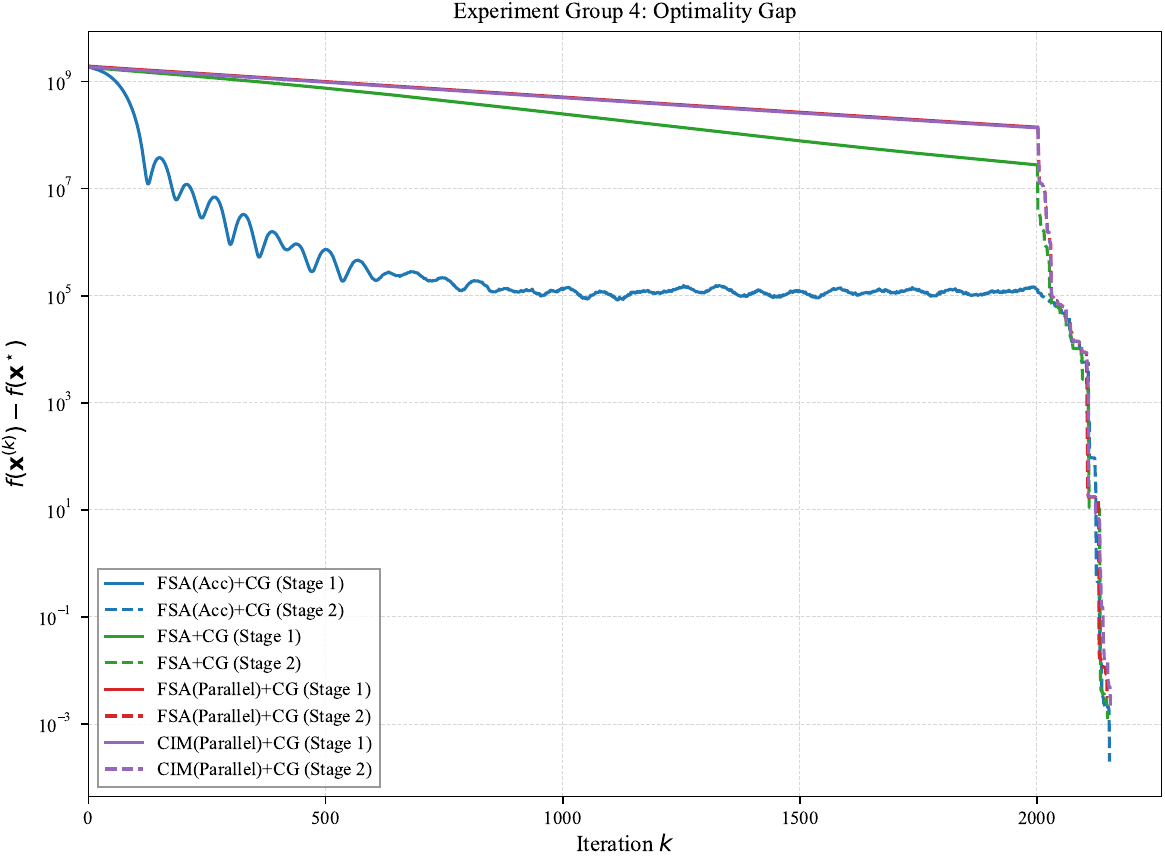}
        \caption{Gap (Group 4)}
    \end{subfigure}
    \caption{Dramatic secondary gradient descent achieved by integrating a Conjugate Gradient continuation phase.}
    \label{fig:exp4_convergence}
\end{figure*}

\begin{figure*}[htbp]
    \centering
    \includegraphics[width=0.7\textwidth]{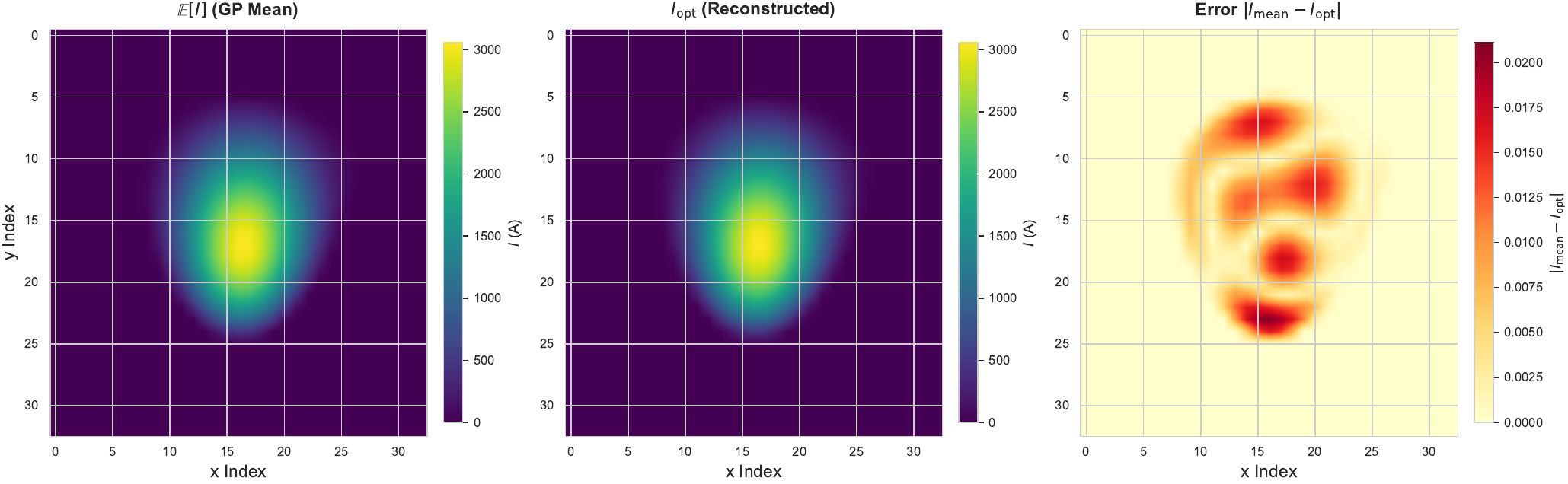}
    \caption{Ultimate high-fidelity current density distribution obtained via CIM parallel computation followed by CG refinement.}
    \label{fig:exp4_current}
\end{figure*}

As depicted in Fig.~\ref{fig:exp4_convergence}, the introduction of the CG phase induced a dramatic secondary gradient descent in the objective gap architecture across all primary methods tested. Most critically, the CIM parallel band-partitioned architecture, initially run for 2,000 iterations, converged smoothly to a near-zero theoretical operational gap upon initialization of the 150-cycle CG correction. The resulting current density distribution (Fig.~\ref{fig:exp4_current}) exhibits an extraordinary degree of fidelity, effectively eliminating numerical artifacts. 

\section{Discussion}\label{sec:discussion}

The experimental results validate that the proposed hybrid BCD-CIM framework represents a major computational advance for real-time plasma diagnostics. In magnetic confinement fusion devices, achieving millisecond-level reconstructions of the internal plasma current density profile is crucial for active feedback control of magnetohydrodynamic (MHD) instabilities, such as neoclassical tearing modes, and for regulating transport barriers. Traditional Bayesian sampling schemes like MCMC are far too slow for real-time control loops, often requiring hours or days. By combining the global search capability of CIMs with the block-wise spatial decomposition of BCD, our paradigm compresses the optimization runtime from hours to the millisecond regime, resolving a decades-long bottleneck in integrated data analysis.

We choose to focus our primary validation on the reconstruction of the plasma current density profile. Because this inverse problem is modeled as a linear-quadratic system, it possesses a strictly convex, quadratic negative log-posterior. The existence of a unique, deterministic global minimum (MAP estimate) allows us to establish a rigorous, analytically verifiable reference baseline. This enables unambiguous validation of the CIM-based solver's convergence and reconstruction accuracy. {\color{black}For more complex joint inferences involving multiple parameters (such as current and electron density profiles), the bilinear coupling introduces non-convex, non-Gaussian posteriors. After binary discretization, the resulting optimization problem takes the form of a Higher-Order Binary Optimization (HOBO) problem, for which the global optimum is challenging to verify.} Evaluating the solver on the quadratic current reconstruction thus provides a clean and mathematically sound proof-of-principle verification of our hybrid solver's capabilities before extending it to non-convex regimes.

A critical finding in our work is the impact of spatial partitioning topology on convergence. Our systematic comparisons between banded partitions and 2D block-wise partitions ($4 \times 4$ and $5 \times 5$ grids) reveal that compact 2D block partitioning yields significantly faster and more stable convergence. From an optimization perspective, local square blocks isolate spatially localized physical dependencies, reducing the strength of coupling terms across block boundaries. This effectively minimizes cross-block coordination overhead during iterations and prevents long-range error propagation. The $4 \times 4$ block partition configuration, accelerated by FISTA momentum, represents the optimal design point for balancing intra-block optimization complexity and inter-block communication.

{\color{black}While the current work utilizes BCD to decompose the high-dimensional grid into manageable blocks to accommodate hardware limits, the ultimate limit of this technology lies in the direct, one-step optimization of the entire plasma grid on ultra-large-scale CIM hardware. With the rapid development of integrated photonics and spatial light modulator-based CIMs, hardware spin capacities are projected to reach $10^5$ to $10^6$ spins in the near future. Under such ultra-large-scale architectures, the entire current density profile can be mapped onto a single, global Ising Hamiltonian. This would bypass the sequential coordination overhead of BCD, allowing the system to converge to the global MAP estimate in a single physical bifurcation sweep (typically under $100\,\mu\text{s}$), thereby providing a definitive solution for real-time MHD control and profile regulation in future fusion reactors.}

{\color{black}Furthermore, the two-stage hybrid optimization paradigm resolves the inherent precision limits of discrete analog Ising solvers. While binary encoding allows us to map the continuous parameters to an Ising Hamiltonian suitable for CIM hardware, it introduces quantization errors that manifest as high-frequency spatial ripples in the reconstructed profile. The terminal Conjugate Gradient (CG) phase uses the continuous gradient of the posterior energy landscape, starting from the near-optimal basin located by the CIM global sweep. Because the CIM phase successfully avoids local minima traps and places the state near the global optimum, the CG refinement requires fewer than 150 iterations to achieve sub-percent accuracy. This hybrid coherent-classical paradigm leverages the global search power of analog physical systems and the local convergence precision of continuous numerical algorithms, establishing a new computational benchmark for structural inverse problems in fusion physics.}

\section{Methods}\label{sec:methods}

\subsection{Bayesian Formulation and Diagnostic Forward Models}

The reconstruction of the plasma current density profile is formulated as a high-dimensional linear-quadratic inverse problem. We represent the discretized toroidal current density over a spatial grid of size $N$ as a vector $\mathbf{I} = (I_1, I_2, \dots, I_N)^\top \in \mathbb{R}^N$. Our objective is to infer the current distribution $\mathbf{I}$ from a set of magnetic diagnostic measurements $\mathbf{D}_{\mathrm{Mag}} \in \mathbb{R}^M$ (comprising flux loops and magnetic probes). 

{\color{black}We focus our primary validation on this single-parameter reconstruction because it yields a quadratic negative log-posterior with a mathematically deterministic global minimum (MAP estimate). This deterministic property provides a rigorous reference baseline to evaluate the convergence accuracy of our coherent analog solver. For generalizability, in Section~\ref{sec:multi_diag} we also detail the multi-diagnostic joint inference framework where non-Gaussian bilinear coupling yields a non-convex posterior, which demonstrates the general applicability of our mapping framework to complex systems.}

Using Bayes' theorem, the posterior probability distribution of the current profile is given by:
\begin{equation}
\begin{aligned}
    P(\mathbf{I} \mid \mathbf{D}_{\mathrm{Mag}}) &= \frac{P(\mathbf{I}) P(\mathbf{D}_{\mathrm{Mag}} \mid \mathbf{I})}{P(\mathbf{D}_{\mathrm{Mag}})} \\
    &\propto P(\mathbf{I}) P(\mathbf{D}_{\mathrm{Mag}} \mid \mathbf{I}),
\end{aligned}
\end{equation}
where $P(\mathbf{I})$ represents the prior distribution encoding physical constraints, and $P(\mathbf{D}_{\mathrm{Mag}} \mid \mathbf{I})$ is the likelihood function representing the diagnostic data fidelity.

\subsubsection{Diagnostic Likelihood Model}
The physical forward model relating the internal plasma current distribution to the external magnetic measurements is derived from the Biot-Savart law and Green's functions. Under the assumption of axisymmetric toroidal geometry, the forward model is linear:
\begin{equation}
    \mathbf{D}_{\mathrm{Mag}} = \mathbf{R} \mathbf{I} + \mathbf{C} + \boldsymbol{\epsilon},
\end{equation}
where $\mathbf{R} \in \mathbb{R}^{M \times N}$ is the response matrix, $\mathbf{C} \in \mathbb{R}^M$ represents background magnetic fields (e.g., from poloidal field coils) and sensor offsets, and $\boldsymbol{\epsilon} \sim \mathcal{N}(\mathbf{0}, \Sigma_{\mathrm{Mag}})$ is the measurement noise, modeled as zero-mean multivariate Gaussian noise with covariance matrix $\Sigma_{\mathrm{Mag}}$. The diagnostic likelihood function is thus expressed as:
\begin{equation}
\begin{aligned}
    P(\mathbf{D}_{\mathrm{Mag}} \mid \mathbf{I}) \propto & \exp \left( -\frac{1}{2} (\mathbf{R}\mathbf{I} + \mathbf{C} - \mathbf{D}_{\mathrm{Mag}})^\top \right. \\
    & \left. \times \Sigma_{\mathrm{Mag}}^{-1} (\mathbf{R}\mathbf{I} + \mathbf{C} - \mathbf{D}_{\mathrm{Mag}}) \right).
\end{aligned}
\end{equation}

\subsubsection{Conditional Autoregressive (CAR) Prior Model}
To address the ill-posedness of the magnetic reconstruction, we enforce spatial smoothness constraints via a Conditional Autoregressive (CAR) prior. Under the CAR framework, the prior distribution of the current at a given grid node $i$, conditioned on its neighbors $\mathcal{N}(i)$, is Gaussian:
\begin{equation}
    I_i \mid \mathbf{I}_{-i} \sim \mathcal{N} \left( \alpha \sum_{j \in \mathcal{N}(i)} \frac{W_{ij}}{D_{W,ii}} I_j, \frac{\tau^2}{D_{W,ii}} \right),
\end{equation}
where $\mathbf{W}$ is the spatial adjacency matrix ($W_{ij} = 1$ if node $i$ and $j$ are adjacent, and $0$ otherwise), $\mathbf{D}_W = \mathrm{diag}(D_{W,11}, \dots, D_{W,NN})$ is the diagonal degree matrix with $D_{W,ii} = \sum_{j} W_{ij}$, $\alpha \in [0, 1)$ controls the strength of spatial association, and $\tau^2$ is a scale parameter. The joint prior distribution over the entire grid is given by:
\begin{equation}
\begin{aligned}
    P(\mathbf{I}) &\propto \exp \left( -\frac{1}{2\tau^2} \mathbf{I}^\top (\mathbf{D}_W - \alpha \mathbf{W}) \mathbf{I} \right) \\
    &= \exp \left( -\frac{1}{2} \mathbf{I}^\top \Sigma_I^{-1} \mathbf{I} \right),
\end{aligned}
\end{equation}
where the prior precision matrix is defined as $\Sigma_I^{-1} = \frac{1}{\tau^2} (\mathbf{D}_W - \alpha \mathbf{W})$.

\subsubsection{Posterior Energy Minimization}
We obtain the maximum a posteriori (MAP) estimation of the current profile by minimizing the negative log-posterior, which serves as the energy function:
\begin{equation}
\begin{aligned}
    \mathcal{L}(\mathbf{I}) = & \frac{1}{2} \mathbf{I}^\top \Sigma_I^{-1} \mathbf{I} \\
    & + \frac{1}{2} (\mathbf{R}\mathbf{I} + \mathbf{C} - \mathbf{D}_{\mathrm{Mag}})^\top \Sigma_{\mathrm{Mag}}^{-1} \\
    & \quad \times (\mathbf{R}\mathbf{I} + \mathbf{C} - \mathbf{D}_{\mathrm{Mag}}).
\end{aligned}
\end{equation}
Expanding this quadratic form yields:
\begin{equation}
    \mathcal{L}(\mathbf{I}) = \frac{1}{2} \mathbf{I}^\top \mathbf{Q} \mathbf{I} + \mathbf{b}^\top \mathbf{I} + \text{const},
\end{equation}
where the global Hessian matrix $\mathbf{Q}$ and the linear coefficient vector $\mathbf{b}$ are:
\begin{align}
    \mathbf{Q} &= \Sigma_I^{-1} + \mathbf{R}^\top \Sigma_{\mathrm{Mag}}^{-1} \mathbf{R}, \\
    \mathbf{b} &= \mathbf{R}^\top \Sigma_{\mathrm{Mag}}^{-1} (\mathbf{C} - \mathbf{D}_{\mathrm{Mag}}).
\end{align}

\subsection{Multi-Diagnostic Joint Inference and Non-Gaussian Coupling}\label{sec:multi_diag}

In integrated data analysis (IDA), joint inference of multiple plasma parameters is conducted to resolve physical inconsistencies across diagnostics. We consider the joint estimation of the current density profile $\mathbf{I}$ and the electron density profile $\mathbf{n}_e \in \mathbb{R}^N$ using magnetic coils, interferometry, and polarimetry (Faraday rotation). The joint prior is factorized as $P(\mathbf{I}, \mathbf{n}_e) = P(\mathbf{I}) P(\mathbf{n}_e)$. Assuming conditional independence among measurements, the joint likelihood is:
\begin{equation}
\begin{aligned}
    P(\mathbf{D}_{\mathrm{all}} \mid \mathbf{I}, \mathbf{n}_e) = & P(\mathbf{D}_{\mathrm{Mag}} \mid \mathbf{I}) P(\mathbf{D}_{\mathrm{HCN}} \mid \mathbf{n}_e) \\
    & \times P(\mathbf{D}_{\psi} \mid \mathbf{I}, \mathbf{n}_e),
\end{aligned}
\end{equation}
where $\mathbf{D}_{\mathrm{HCN}}$ represents interferometer line-integrated density, and $\mathbf{D}_{\psi}$ represents polarimeter Faraday rotation angles.

The polarimetry forward model exhibits a bilinear coupling of the electron density and parallel magnetic field:
\begin{equation}
    \theta_{\psi, i} \propto \int_{\mathcal{L}_i} n_e B_{\parallel} dl \approx \sum_{j \in \mathcal{L}_i} n_{e,j} B_{\parallel, j} \Delta l.
\end{equation}
Since the local magnetic field is linearly related to the plasma current, $B_{\parallel, j} = (\mathbf{R}_{\parallel} \mathbf{I})_j + C_{\parallel, j}$, the discretized polarimetry forward model can be written in matrix form as:
\begin{equation}
    \mathbf{D}_{\psi} = \mathbf{R}_{\psi} \mathbf{N}_e (\mathbf{R}_{\parallel} \mathbf{I} + \mathbf{C}_{\parallel}) + \boldsymbol{\epsilon}_{\psi},
\end{equation}
where $\mathbf{N}_e = \mathrm{diag}(\mathbf{n}_e)$ is the diagonal electron density matrix and $\boldsymbol{\epsilon}_{\psi} \sim \mathcal{N}(\mathbf{0}, \Sigma_{\psi})$. The dependence of the polarimetric response on the product of density and current introduces a multiplicative coupling term $\mathbf{n}_e^\top \mathbf{T} \mathbf{I}$ in the forward model, rendering the posterior distribution non-Gaussian. The negative log-posterior is a higher-order polynomial:
\begin{equation}
\begin{aligned}
    \mathcal{L}(\mathbf{I}, \mathbf{n}_e) &= -\log P(\mathbf{I}, \mathbf{n}_e \mid \mathbf{D}_{\mathrm{all}}) \\
    &= \mathcal{L}_0(\mathbf{I}) + \mathcal{L}_1(\mathbf{n}_e) \\
    & \quad + \frac{1}{2} \left\| \mathbf{R}_{\psi} \mathbf{N}_e (\mathbf{R}_{\parallel} \mathbf{I} + \mathbf{C}_{\parallel}) - \mathbf{D}_{\psi} \right\|_{\Sigma_{\psi}^{-1}}^2,
\end{aligned}
\end{equation}
which is a non-convex fourth-order polynomial optimization problem. Minimizing this objective corresponds to a Higher-order Binary Optimization (HOBO) problem upon binary discretization.

To solve this on quadratic analog devices like coherent Ising machines, we perform quadratization by introducing auxiliary variables. For any cubic or quartic product terms $z_1 z_2 z_3$, we introduce an auxiliary binary variable $y_{12} \in \{0, 1\}$ to represent the product $z_1 z_2$, and append a penalty function to the objective:
\begin{equation}
\begin{aligned}
    H_{\mathrm{penalty}}(z_1, & z_2, y_{12}) = P \cdot (z_1 z_2 - 2 z_1 y_{12} \\
    & - 2 z_2 y_{12} + 3 y_{12}),
\end{aligned}
\end{equation}
where $P > 0$ is a penalty coefficient. Minimizing the penalty enforces $y_{12} = z_1 z_2$ at the ground state, reducing the degree of the polynomial to quadratic (QUBO).

\subsection{Continuous-to-Binary Mapping and Adaptive Binarization}

{\color{black}The conversion of continuous physical models (via discretization and binary expansion) to discrete Quadratic Unconstrained Binary Optimization (QUBO) formulations is a general-purpose methodology. It provides a universal mathematical interface that bridges continuous physical fields and discrete analog Ising solvers (such as coherent Ising machines or quantum annealers). This formulation is not restricted to tokamak profile reconstructions; it is widely applicable to a broad spectrum of continuous inverse problems in physics and engineering, including geophysical tomography, acoustic wave inversion, and structural health monitoring. Any continuous optimization problem with a polynomial energy landscape can be mapped onto analog spin networks through appropriate discretization, scaling, and penalty-based quadratization techniques.}

To map the continuous optimization problem to binary spin architectures, each continuous variable $I_i$ is encoded using a finite binary representation. A naive binary encoding maps a continuous variable to $K$ bits:
\begin{equation}
    I_i = I_{\min, i} + \Delta I_i \sum_{k=1}^{K} 2^{-k} x_{i,k}, \quad x_{i,k} \in \{0,1\}.
\end{equation}
In vector notation, we write:
\begin{equation}
    \mathbf{I} = \mathbf{B} \mathbf{x} + \mathbf{I}_{\min},
\end{equation}
where $\mathbf{x} \in \{0,1\}^{NK}$ is the concatenated binary vector, and $\mathbf{B} \in \mathbb{R}^{N \times NK}$ is the block-diagonal encoding matrix.

\subsubsection{High Bit-Cost Dilemma and Two-Stage Adaptive Binarization}\label{subsubsec:two_stage_binarization}
Achieving high numerical precision requires large bit depths $K$. However, this causes the size of the binary vector $\mathbf{x}$ to scale as $NK$. For large-scale grids (e.g., $N \approx 10^3$), a high bit depth (e.g., $K=16$) yields $1.6 \times 10^4$ spins, which exceeds the \textcolor{black}{spin capacity and stability limits of current analog Ising solvers and optical hardware}. 

To overcome this dilemma, we propose a Two-Stage Adaptive Binarization (coarse-to-fine) scheme:
\begin{enumerate}
    \item \textbf{Coarse Anchoring Stage}: The optimization is solved over the full physical range $[I_{\min}, I_{\max}]$ using a coarse resolution with small bit depth $K_{\mathrm{coarse}}$ (e.g., $K=4$). This global search yields a coarse MAP profile $\mathbf{I}^{(0)}$.
    \item \textbf{Fine-Grained Refining Stage}: The search range for each variable is dynamically zoomed in around the coarse estimate:
    \begin{equation}
        I_i \in [I_i^{(0)} - \delta_i, I_i^{(0)} + \delta_i],
    \end{equation}
    where $\delta_i$ is a local scale factor. We then re-encode $I_i$ over this narrow interval using a fine bit depth $K_{\mathrm{fine}}$ (e.g., $K=4$).
\end{enumerate}
This adaptive zooming strategy reduces the active spin variables to $N K_{\mathrm{coarse}}$ and $N K_{\mathrm{fine}}$, dramatically compressing the \textcolor{blue}{spin} overhead while retaining a high effective resolution equivalent to $K = 16$.

\subsubsection{QUBO Formulation}
Substituting $\mathbf{I} = \mathbf{B} \mathbf{x} + \mathbf{I}_{\min}$ into the continuous energy function $\mathcal{L}(\mathbf{I})$ yields:
\begin{equation}
\begin{aligned}
    \mathcal{L}(\mathbf{x}) = & \frac{1}{2} (\mathbf{B}\mathbf{x} + \mathbf{I}_{\min})^\top \mathbf{Q} (\mathbf{B}\mathbf{x} + \mathbf{I}_{\min}) \\
    & + \mathbf{b}^\top (\mathbf{B}\mathbf{x} + \mathbf{I}_{\min}).
\end{aligned}
\end{equation}
Expanding and dropping constant terms yields the standard QUBO problem:
\begin{equation}
    \min_{\mathbf{x} \in \{0,1\}^{NK}} \mathbf{x}^\top \mathbf{Q}_{\mathrm{QUBO}} \mathbf{x},
\end{equation}
where the QUBO matrix $\mathbf{Q}_{\mathrm{QUBO}} \in \mathbb{R}^{NK \times NK}$ is constructed as:
\begin{equation}
    \mathbf{Q}_{\mathrm{QUBO}} = \frac{1}{2} \mathbf{B}^\top \mathbf{Q} \mathbf{B} + \mathrm{diag}\left( \mathbf{B}^\top ( \mathbf{Q} \mathbf{I}_{\min} + \mathbf{b} ) \right).
\end{equation}
Here, the linear term is folded into the diagonal of $\mathbf{Q}_{\mathrm{QUBO}}$ leveraging the binary identity $x_i^2 = x_i$.

\subsubsection{\color{black}Generalization to Non-Quadratic Inference via Taylor Approximation}

{\color{black}For general parameter inference problems where the negative log-posterior (or the physical objective function) $\mathcal{L}(\boldsymbol{\theta})$ is non-linear and non-quadratic with respect to the continuous parameters $\boldsymbol{\theta} \in \mathbb{R}^d$, we can resolve this complexity by employing a local second-order Taylor expansion. Specifically, around a reference parameter state $\boldsymbol{\theta}^{(t)}$ at iteration step $t$, the objective function is approximated as:
\begin{equation}
\begin{aligned}
    \mathcal{L}(\boldsymbol{\theta}) \approx & \mathcal{L}(\boldsymbol{\theta}^{(t)}) + \nabla \mathcal{L}(\boldsymbol{\theta}^{(t)})^\top (\boldsymbol{\theta} - \boldsymbol{\theta}^{(t)}) \\
    & + \frac{1}{2} (\boldsymbol{\theta} - \boldsymbol{\theta}^{(t)})^\top \mathbf{H}(\boldsymbol{\theta}^{(t)}) (\boldsymbol{\theta} - \boldsymbol{\theta}^{(t)}),
\end{aligned}
\end{equation}
where $\nabla \mathcal{L}(\boldsymbol{\theta}^{(t)}) \in \mathbb{R}^d$ and $\mathbf{H}(\boldsymbol{\theta}^{(t)}) \in \mathbb{R}^{d \times d}$ denote the gradient vector and the Hessian matrix of the objective function evaluated at $\boldsymbol{\theta}^{(t)}$, respectively. By defining the parameter update vector $\Delta \boldsymbol{\theta} = \boldsymbol{\theta} - \boldsymbol{\theta}^{(t)}$, the optimization problem is reduced locally to a quadratic minimization form:
\begin{equation}
    \min_{\Delta \boldsymbol{\theta}} \frac{1}{2} \Delta \boldsymbol{\theta}^\top \mathbf{H}(\boldsymbol{\theta}^{(t)}) \Delta \boldsymbol{\theta} + \nabla \mathcal{L}(\boldsymbol{\theta}^{(t)})^\top \Delta \boldsymbol{\theta}.
\end{equation}
This local quadratic approximation can be mapped onto the binary spin space of the Coherent Ising Machine (CIM) by discretizing the step $\Delta \boldsymbol{\theta}$ using the encoding matrix $\mathbf{B}$ as $\Delta \boldsymbol{\theta} = \mathbf{B} \mathbf{x} + \Delta \boldsymbol{\theta}_{\min}$, where $\mathbf{x} \in \{0, 1\}^{dK}$. Substituting this encoding yields a standard quadratic unconstrained binary optimization (QUBO) problem:
\begin{equation}
    \min_{\mathbf{x} \in \{0, 1\}^{dK}} \mathbf{x}^\top \mathbf{Q}_{\mathrm{QUBO}}^{(t)} \mathbf{x},
\end{equation}
where the QUBO matrix is defined as:
\begin{equation}
\begin{aligned}
    \mathbf{Q}_{\mathrm{QUBO}}^{(t)} = & \frac{1}{2} \mathbf{B}^\top \mathbf{H}(\boldsymbol{\theta}^{(t)}) \mathbf{B} \\
    & + \mathrm{diag}\Big( \mathbf{B}^\top \big( \mathbf{H}(\boldsymbol{\theta}^{(t)}) \Delta \boldsymbol{\theta}_{\min} \\
    & + \nabla \mathcal{L}(\boldsymbol{\theta}^{(t)}) \big) \Big).
\end{aligned}
\end{equation}
Solving this QUBO on the CIM yields the optimal binary update step $\mathbf{x}^*$, which determines the updated parameters: $\boldsymbol{\theta}^{(t+1)} = \boldsymbol{\theta}^{(t)} + \Delta \boldsymbol{\theta}^*$. By iteratively constructing and solving these local quadratic Taylor approximations, CIM-based optimization can be generalized to solve arbitrary non-linear and non-quadratic continuous parameter inference problems.}

\subsection{Ising Mapping and Coherent Ising Machine Dynamics}

The binary variables $\mathbf{x} \in \{0, 1\}^{NK}$ are mapped to Ising spins $\mathbf{s} \in \{-1, 1\}^{NK}$ using the coordinate transformation $x_i = \frac{1+s_i}{2}$. The QUBO objective is transformed into the Ising spin Hamiltonian (omitting constant energy offsets):
\begin{equation}
    E(\mathbf{s}) = \sum_{i < j} J_{ij} s_i s_j + \sum_i h_i s_i,
\end{equation}
where the coupling matrix $J_{ij}$ and the local field vector $h_i$ are given by:
\begin{align}
    J_{ij} &= \frac{1}{4} (Q_{\mathrm{QUBO}})_{ij}, \quad (i \neq j), \\
    h_i &= \frac{1}{2} (Q_{\mathrm{QUBO}})_{ii} + \frac{1}{2} \sum_{j \neq i} (Q_{\mathrm{QUBO}})_{ij}.
\end{align}

A Coherent Ising Machine (CIM) physically solves this spin Hamiltonian by utilizing a network of mutually coupled degenerate optical parametric oscillators (DOPOs). In the classical limit, the in-phase amplitudes $x_i$ of the optical pulses circulating in the fiber ring cavity evolve according to the following set of Stochastic Differential Equations (SDEs):
\begin{equation}
\begin{aligned}
    d x_i = & \left( (p - 1 - x_i^2) x_i + \xi \sum_{j \neq i} J_{ij} x_j \right) dt \\
    & + \sqrt{\frac{g}{2} (x_i^2 + 1)} dW_i,
\end{aligned}
\end{equation}
where $p$ is the pump amplitude (gain parameter) representing energy injection into the cavity, $\xi$ is the coupling coefficient representing the feedback strength, $g$ represents the quantum noise level, and $dW_i$ is a standard independent Wiener process representing quantum vacuum fluctuations. During the computation, the pump parameter $p(t)$ is ramped slowly from below OPO threshold ($p < 1$) to above threshold ($p > 1$), causing the system to undergo a pitchfork bifurcation. Gain competition forces the optical pulses to lock into the phase state representing the global minimum of the Ising energy landscape.

\subsection{Domain-Decomposed Block Coordinate Descent (BCD-CIM) Framework}

To scale the algorithm to large grids exceeding the physical spin limits of the quantum hardware, we employ a domain-decomposed Block Coordinate Descent (BCD) framework~\cite{tseng2001convergence, wright2015coordinate}. The spatial grid of size $N$ is partitioned into $m$ disjoint subdomains:
\begin{equation}
    \bigcup_{i=1}^m \Omega_i = \{1, 2, \dots, N\}, \quad \Omega_i \cap \Omega_j = \emptyset \ (i \neq j).
\end{equation}
Let $\mathbf{I}_i \in \mathbb{R}^{|\Omega_i|}$ represent the current subvector for subdomain $\Omega_i$, and $\mathbf{I}_{-i}$ denote the variables in all other subdomains. In each outer iteration step, we update the subdomains sequentially. 

The global quadratic objective function is:
\begin{equation}
    F(\mathbf{I}) = \frac{1}{2} \mathbf{I}^\top \mathbf{Q} \mathbf{I} + \mathbf{q}^\top \mathbf{I}.
\end{equation}
We partition the Hessian matrix $\mathbf{Q}$ and vector $\mathbf{q}$ block-wise with respect to the subdomain indices:
\begin{equation}
    \mathbf{Q} = \begin{pmatrix} \mathbf{Q}_{ii} & \mathbf{Q}_{i,-i} \\ \mathbf{Q}_{-i,i} & \mathbf{Q}_{-i,-i} \end{pmatrix}, \quad \mathbf{q} = \begin{pmatrix} \mathbf{q}_i \\ \mathbf{q}_{-i} \end{pmatrix}.
\end{equation}
Fixing the variables in all other subdomains $\mathbf{I}_{-i}$, the local subproblem for block $i$ is:
\begin{equation}
\begin{aligned}
    F_i(\mathbf{I}_i \mid \mathbf{I}_{-i}) = & \frac{1}{2} \mathbf{I}_i^\top \mathbf{Q}_{ii} \mathbf{I}_i \\
    & + \left( \mathbf{Q}_{i,-i} \mathbf{I}_{-i} + \mathbf{q}_i \right)^\top \mathbf{I}_i + \mathrm{const},
\end{aligned}
\end{equation}
which is equivalent to a local quadratic optimization problem with Hessian $\mathbf{Q}_{ii}$ and an effective local linear coefficient:
\begin{equation}
    \mathbf{h}_i^{\mathrm{eff}} = \mathbf{Q}_{i,-i} \mathbf{I}_{-i} + \mathbf{q}_i.
\end{equation}
The local linear coefficient $\mathbf{h}_i^{\mathrm{eff}}$ incorporates the physical boundary conditions and the inductive couplings from all external current elements. This local subproblem is converted to a QUBO of size $|\Omega_i| K$ and solved on the CIM.

We evaluate two geometric partitioning topologies:
\begin{itemize}
    \item \textbf{Banded Partitions}: The grid is sliced into 1D horizontal or vertical strips of bandwidth $W$.
    \item \textbf{Square Block Partitions}: The grid is divided into 2D subgrids (e.g., $4 \times 4$ and $5 \times 5$ layouts).
\end{itemize}
Square block partitions isolate local physical correlations more effectively than 1D strips, minimizing the cross-boundary coupling energy ($\|\mathbf{Q}_{i,-i}\|$) and leading to faster and more stable BCD convergence.

\subsection{Preconditioning for Ill-Conditioned Systems}

The Hessian matrix $\mathbf{Q}$ for the plasma current reconstruction is highly ill-conditioned, with a condition number $\kappa(\mathbf{Q}) \approx 3.12 \times 10^7$, due to the decaying nature of magnetic Green's functions and the spatial derivative terms in the CAR prior. This ill-conditioning causes BCD iterations to stall and oscillate.

To accelerate convergence, we apply a preconditioning coordinate transformation. We compute an incomplete Cholesky factorization of the Hessian:
\begin{equation}
    \mathbf{Q} \approx \mathbf{L} \mathbf{L}^\top,
\end{equation}
and define the preconditioning matrix as $\mathbf{P} = \mathbf{L}$. We then transform the variables from physical space $\mathbf{I}$ to preconditioned space $\mathbf{y}$ via:
\begin{equation}
    \mathbf{I} = \mathbf{P}^{-\top} \mathbf{y}.
\end{equation}
Substituting this transformation into the global objective yields:
\begin{equation}
\begin{aligned}
    F(\mathbf{y}) &= \frac{1}{2} \mathbf{y}^\top \left( \mathbf{P}^{-1} \mathbf{Q} \mathbf{P}^{-\top} \right) \mathbf{y} + \left( \mathbf{P}^{-1} \mathbf{q} \right)^\top \mathbf{y} \\
    &= \frac{1}{2} \mathbf{y}^\top \mathbf{A} \mathbf{y} + \mathbf{b}_y^\top \mathbf{y},
\end{aligned}
\end{equation}
where the preconditioned Hessian $\mathbf{A} = \mathbf{P}^{-1} \mathbf{Q} \mathbf{P}^{-\top}$ has a condition number of $\kappa(\mathbf{A}) \approx 1.63$. The BCD-CIM algorithm is executed in the preconditioned space $\mathbf{y}$, which features nearly circular energy contours, enabling rapid and stable convergence. The physical current profile is recovered via $\mathbf{I} = \mathbf{P}^{-\top} \mathbf{y}$ after convergence.

\subsection{Nesterov Acceleration (FISTA) and Convergence Criteria}

To accelerate the propagation of physical information across subdomain interfaces, we integrate a Nesterov-type momentum scheme (FISTA)~\cite{beck2009fast} into the BCD coordinate updates. Let $\mathbf{y}^{(k)}$ represent the preconditioned state at iteration $k$. The update steps are formulated as:
\begin{align}
    \mathbf{z}^{(k)} &= \mathbf{y}^{(k)} + \beta^{(k)} \left( \mathbf{y}^{(k)} - \mathbf{y}^{(k-1)} \right), \\
    \mathbf{y}^{(k+1)}_i &= \arg\min_{\mathbf{x}} F_i\left(\mathbf{x} \mid \mathbf{z}^{(k)}_{-i}\right), \quad i = 1, \dots, m,
\end{align}
where the momentum coefficient $\beta^{(k)}$ is updated at each step as:
\begin{align}
    \beta^{(k)} &= \frac{t^{(k)} - 1}{t^{(k+1)}}, \\
    t^{(k+1)} &= \frac{1 + \sqrt{1 + 4 (t^{(k)})^2}}{2}, \quad t^{(0)} = 1.
\end{align}

The outer BCD loop terminates when the following convergence criteria are satisfied:
\begin{align}
    \left\| \mathbf{y}^{(k+1)} - \mathbf{y}^{(k)} \right\|_{\infty} &\le \varepsilon, \\
    \left| F(\mathbf{y}^{(k+1)}) - F(\mathbf{y}^{(k)}) \right| &\le \delta,
\end{align}
where $\varepsilon$ and $\delta$ are user-defined convergence tolerances, or when the iteration count reaches $K_{\max}$.

\subsection{Two-Stage Hybrid Optimization Strategy}

Binary discretization of continuous profiles introduces quantization errors, which manifest as high-frequency spatial ripples on the reconstructed profile near the convergence floor. To achieve scientific-grade reconstruction accuracy while maintaining high computational speed, we adopt a two-stage hybrid optimization strategy:
\begin{itemize}
    \item \textbf{Stage 1: Global BCD-CIM Search}: We execute the preconditioned BCD-CIM algorithm (with FISTA acceleration) for 2,000 iterations. This analog-accelerated phase rapidly explores the global parameter space, bypasses local minima traps, and brings the state to the immediate vicinity of the global optimum.
    \item \textbf{Stage 2: Continuous Conjugate Gradient Refinement}: The state obtained from Stage 1 is used as the initial guess $\mathbf{I}^{(0)}_{\mathrm{CG}}$ for a continuous Conjugate Gradient (CG) solver. Working in the continuous domain, the CG solver performs local refinement along the exact gradient of the posterior energy landscape:
    \begin{equation}
        \mathbf{g} = \nabla \mathcal{L}(\mathbf{I}) = \mathbf{Q} \mathbf{I} + \mathbf{b}.
    \end{equation}
    Because the starting point is already in the global basin, the CG solver converges to the exact MAP solution in fewer than 150 iterations, eliminating quantization ripples and reaching the precision floor of the double-precision floating-point format.
\end{itemize}
This hybrid paradigm effectively combines the global optimization strengths of discrete analog solvers with the high-precision refinement capabilities of classical continuous optimization.



\section{Competing interests}
The authors declare no competing interests.

\bibliographystyle{sn-nature}
\bibliography{references}

\end{document}